\documentclass[%
 aip,
 amsmath,amssymb,
 reprint,%
]{revtex4-1}

\usepackage{graphicx}
\usepackage{dcolumn}
\usepackage{bm}
\usepackage{amsfonts}
\usepackage{amsmath} 
\usepackage{physics} 

\usepackage[utf8]{inputenc}
\usepackage[T1]{fontenc}
\usepackage{mathptmx}
\usepackage{etoolbox}

\makeatletter
\def\@email#1#2{%
 \endgroup
 \patchcmd{\titleblock@produce}
  {\frontmatter@RRAPformat}
  {\frontmatter@RRAPformat{\produce@RRAP{*#1\href{mailto:#2}{#2}}}\frontmatter@RRAPformat}
  {}{}
}%
\makeatother

\graphicspath{{./diagram/}} 

\usepackage{xcolor}
\usepackage{xr}
\usepackage{siunitx}

\DeclareSIUnit[number-unit-product = {\,}]\cal{cal}
\DeclareSIUnit{\kcal}{\kilo\cal}

\newcommand{\kB}{k_\mathrm{B}}

\newcommand{\Rc}{R_\mathrm{c}}
\newcommand{\Rs}{R_\mathrm{s}}
\newcommand{\gc}{g_\mathrm{c}}
\newcommand{\sigmac}{\sigma_\mathrm{c}}

\newcommand{\FFGI}{\bm{F}_{\mathrm{FG},I}}
\newcommand{\tauFGI}{\bm{\tau}_{\mathrm{FG},I}}
\newcommand{\WFGI}{\bar{W}_{\mathrm{FG},I}}
\newcommand{\Linst}{L_\mathrm{inst}}

\begin{document}


\title{Anisotropic molecular coarse-graining by force and torque matching with neural networks}
\author{Marltan O. Wilson}
 \affiliation{Department of Chemistry, School of Physics, Chemistry and Earth Sciences, The University of Adelaide, Adelaide, South Australia 5005, Australia}
\author{David M. Huang}%
 \affiliation{Department of Chemistry, School of Physics, Chemistry and Earth Sciences, The University of Adelaide, Adelaide, South Australia 5005, Australia}
 \email{david.huang@adelaide.edu.au}

\date{\today}

\begin{abstract}
We develop a machine-learning method for coarse-graining condensed-phase molecular systems using anisotropic particles. The method extends currently available high-dimensional neural network potentials by addressing molecular anisotropy. We demonstrate the flexibility of the method by parametrizing single-site coarse-grained models of a rigid small molecule (benzene) and a semi-flexible organic semiconductor (sexithiophene), attaining structural accuracy close to the all-atom models for both molecules at considerably lower computational expense. The machine-learning method of constructing the coarse-grained potential is shown to be straightforward and sufficiently robust to capture anisotropic interactions and many-body effects. The method is validated through its ability to reproduce the structural properties of the small molecule's liquid phase and the phase transitions of the semi-flexible molecule over a wide temperature range.
\end{abstract}

\maketitle

\section{Introduction\label{sec:intro}}

Machine learning is quickly becoming an invaluable tool in the search, analysis, and development of new materials.\cite{butler2018machine,moosavi2020role}
Neural networks, in particular, have had major recent success in areas ranging from predicting the folded conformation of biological macromolecules such as proteins \cite{tunyasuvunakool2021} to developing highly accurate temperature-transferable interatomic potentials.\cite{rowe2020accurate,stocker2022robust} The latter is an important advance in the field of molecular dynamics (MD) simulations.  Improvements in these machine-learning models aim to expand the length and time scale of simulations without sacrificing accuracy.\cite{friederich2021machine,noe2020machine}
Currently used ab initio molecular dynamics simulation models are generally accurate but are computationally expensive, limiting their ability to probe long time scales.\cite{Guo2022,Marx2009} However, machine-learning potentials can produce ab initio accuracy at the computational cost of classical atomistic models. \cite{Wang2018,unke2021machine} 

Even though simulations at the classical MD level are faster than ab initio MD, the speedup is still insufficient to model the large length and time scales needed to fully understand certain phenomena and processes such as supramolecular assembly. It is well known that explicit modeling of high-frequency motion is not critical for describing many phenomena in molecular systems. These simplifications have led to the development of molecular coarse-grained models to study large, complex materials and biological systems.\cite{jin2022bottom} Parameterization of coarse-grained interaction potentials commonly uses one of two strategies: the top-down approach in which parameters are tuned to match macroscopic observables, as exemplified by the Martini model,\cite{marrink2013perspective} and the bottom-up approach in which interactions are derived from the properties of a fine-grained model with more degrees of freedom.\cite{jin2022bottom} By following a similar bottom-up process used to apply machine learning to ab initio MD data, neural-network approaches have been extended to coarse-grained molecular models, further extending the length and time scale of simulations with atomistic accuracy.\cite{Zhang2018,Wang2019}

Neural-network potentials using isotropic coarse-grained particles have several advantages over their pair-wise additive analytical counterparts since they are constructed as many-body potentials. This many-body potential can become costly when multiple coarse-grained particles are needed to preserve the shape anisotropy of a molecular fragment. It is sometimes more accurate and computationally efficient to represent these groups of atoms as a single anisotropic coarse-grained particle such as an ellipsoid, such as in the case of large, rigid, anisotropic molecular fragments. Analytical anisotropic coarse-grained potentials such as the Gay--Berne potential\cite{Gay1981,Berardi1995} were developed to address the poor performance of spherically symmetric potentials in replicating intrinsic anisotropic interactions such as $\pi$-stacking. By modeling rigid anisotropic groups of atoms as ellipsoids, the anisotropic properties of the group are preserved in a single-site model. Shape and interaction anisotropy is especially important for the study of organic semiconductor molecules, which typically consist of highly anisotropic and rigid $\pi$-conjugated units and often form liquid-crystal phases whose morphology strongly affects their performance in devices such as solar cells, transistors, and light-emitting diodes.\cite{boehm2019interplay}

Unlike analytical pair-wise additive potentials such as the Gay--Berne potential, high-dimensional neural-network potentials are constructed based on the immediate neighborhood of a molecule and thus account for many-body effects as well as local density variations. Notable machine-learning implementations of interatomic and intermolecular potentials include the neural-network potentials developed by Behler et al. \cite{Behler2011} The Behler neural-network potentials are constructed from a set of symmetry functions used to represent the invariant properties of the atomic environment of each atom taken from ab initio simulations.  DeepMD \cite{Wang2018} and DeepCG \cite{Zhang2018} are two other neural-network codes constructed for atomistic and coarse-grained simulations, respectively. All these neural-network potentials rely on an invariant representation of the atomic/molecular environment. The CGnets deep-learning approach \cite{Wang2019} employs a prior potential to account for areas in a coarse-grained data set that may not be properly sampled due to high repulsive energies. These interactions are especially important to reproduce the local structure of the simulated material.
 
Machine learning has previously been applied to the parameterization of coarse-grained models with anisotropic particles, \cite{CamposVillalobos2022} but no such implementation has used a nonlinear neural-network optimization method to construct the coarse-grained potential. In this work, we address this gap in knowledge by using a neural network to construct a high-dimensional anisotropic coarse-grained potential. We parameterize the neural-network potential using a recently derived systematic and general bottom-up coarse-graining method called anisotropic force-matching coarse-graining (AFM-CG),\cite{Nguyen2022} which generalizes the multi-scale coarse-graining (MS-CG) method \cite{Noid2008} for isotropic coarse-grained particles to anisotropic particles. The method rigorously accounts for finite-temperature, many-body effects without assuming a specific functional form of the anisotropic coarse-grained potential. It yields general equations relating the forces, torques, masses, and moments of inertia of the coarse-grained particles to properties of a fine-grained (e.g. all-atom) MD simulation based on a mapping between fine-grained and coarse-grained coordinates and momenta, and by matching the equilibrium coarse-grained phase-space distribution with the mapped distribution of the fine-grained system. The previous implementations of the AFM-CG method approximated the coarse-grained potential as a sum of pair interactions between particles.\cite{Nguyen2022} Here, we extend this approach to more general many-body anisotropic interactions described by a neural-network potential. We also extend the approach, which was derived for constant-volume systems in the canonical ensemble to constant-pressure systems by applying a virial-matching condition previously derived for the MS-CG method.\cite{das2010a}

A general coarse-grained potential should capture any temperature-dependent phase transition associated with either melting, annealing, or glass transition temperatures as well as the local structure and density of the material. The focus is on the development of a model for which trained parameters can be easily obtained and one capable of reproducing interaction anisotropy, temperature transferability, and many-body effects. The flexibility of the new model is demonstrated through the matching of structural and thermodynamic properties of condensed-phase systems of a small anisotropic molecule, benzene, and of a larger, more flexible organic semiconductor molecule, sexithiophene.  These two molecules were chosen to determine the conditions under which coarse-grained structural inaccuracy outweighs the computational efficiency of a single-anisotropic-site model.    

\section{Theory}

The key aspects of the theory that underpins the AFM-CG method and its extension to constant pressure via virial matching are summarized below. The reader is referred to Ref.~\citenum{Nguyen2022} for a more detailed description and the full derivation of the method. 
 
The positions $\bm{r}^n =\left\{\bm{r}_1,\bm{r}_2,\ldots,\bm{r}_n\right\}$ of the $n$ fine-grained particles are mapped onto the positions $\bm{R}^N = \left\{\bm{R}_1,\bm{R}_2,\ldots,\bm{R}_N\right\}$ and orientations $\bm{\Omega}^N = \left\{\bm{\Omega}_1,\bm{\Omega}_2,\ldots,\bm{\Omega}_N\right\}$ of the $N$ anisotropic coarse-grained particles. Each fine-grained particle $i$ is mapped to a single coarse-grained particle by defining $N$ non-intersecting subsets, $\zeta_1, \zeta_2, \ldots, \zeta_N$, of the fine-grained particle indices such that $\zeta_I$ contains the indices of the fine-grained particles mapped onto coarse-grained particle $I$. The position $\bm{R}_I$ of coarse-grained particle $I$ is defined to be equal to the center-of-mass of the group of fine-grained particles that are mapped onto it, i.e. 
\begin{equation}
	\bm{R}_I = \frac{\sum_{i \in \zeta_I} {m_i \bm{r}_i}}{\sum_{i \in \zeta_I} m_i} ,
	\label{eq:cgposition}
\end{equation}
where $m_i$ is the mass of fine-grained particle $i$. The orientation 
\begin{equation}
	\bm{\Omega}_I =  
	\begin{bmatrix}
		\bm{\Omega}_{I,1} \\
		\bm{\Omega}_{I,2} \\
		\bm{\Omega}_{I,3}
	\end{bmatrix}
	\label{eq:cgorientation}
\end{equation}
of coarse-grained particle $I$ is specified by the rotation matrix whose components are the particle's three normalized principal axes of inertia, $\bm{\Omega}_{I,q}$ for $q =1,2,3$. These axes are defined to be equal to the corresponding principal axes relative to the center-of-mass of the group of fine-grained particles that are mapped onto the coarse-grained particle. Thus, these axes are the normalized eigenvectors of the inertia tensor
\begin{equation}
	\pmb{\mathbb{I}}_{\mathrm{FG},I} = \displaystyle\sum_{i \in \zeta_I} {m_i (||\Delta \bm{r}_i ||^2 \bm{E} - \Delta \bm{r}_i \Delta \bm{r}_i^\mathrm{T})} , 
\end{equation}
where $\Delta \bm{r}_i = \bm{r}_i - \bm{R}_I$ is the position of fine-grained particle $i$ relative to the center-of-mass (coarse-grained particle position) and $\bm{E}$ is the $3\times3$ identity matrix. From these coordinate mappings and the relationship between generalized coordinates and momenta from Hamilton's equations,\cite{Goldstein2002} mappings from the linear momenta $\bm{p}^n =\left\{\bm{p}_1,\bm{p}_2,\ldots,\bm{p}_n\right\}$ of the fine-grained particles to the linear momenta $\bm{P}^N = \left\{\bm{P}_1,\bm{P}_2,\ldots,\bm{P}_N\right\}$ and angular momenta $\bm{L}^N = \left\{\bm{L}_1,\bm{L}_2,\ldots,\bm{L}_N\right\}$ of the anisotropic coarse-grained particles can also be defined.\cite{Nguyen2022}
The mappings for coarse-grained particle $I$ are
\begin{equation}
	\bm{P}_I = \frac{M_I}{\sum_{i \in \zeta_I} m_i}\sum_{i \in \zeta_I}\bm{p}_i
\end{equation}
and 
\begin{equation}
	\bm{L}_I = \pmb{\mathbb{I}}_I \pmb{\mathbb{I}}_{\mathrm{FG},I}^{-1} \sum_{i \in \zeta_I} \Delta \bm{r}_i \cross \bm{p}_i ,
\end{equation}
respectively, where $\pmb{\mathbb{I}}_I$ is the inertia tensor of coarse-grained particle $I$.
 
Given these mappings, several conditions can be derived that the coarse-grained model must satisfy for its equilibrium coarse-grained phase-space distribution to match the corresponding mapped distribution of the fine-grained system. Consistency between the configuration-space distributions gives the following matching conditions between the forces $\bm{F}_I$ and torques $\bm{\tau}_I$ on coarse-grained particle $I$ and the forces on the fine-grained particles mapped onto it:\cite{Nguyen2022}
\begin{equation} 
	\bm{F}_I(\bm{R}^N,\bm{\Omega}^N) = -\frac{\partial U}{\partial R_I} = \left\langle  \sum_{i \in \zeta_I } {\bm{f}_i} \right\rangle_{R^N,\Omega^N}  
	\label{eq:cgforce}
\end{equation}
and 
\begin{eqnarray} 
	\bm{\tau}_I (\bm{R}^N,\bm{\Omega}^N) = -\sum_{q} \Omega_{I,q}\times \frac{\partial U}{\partial \Omega_{I,q}} = \left\langle \sum_{i \in \zeta_I } {\Delta \bm{r}_i \cross\bm{f}_i} \right\rangle_{R^N,\Omega^N},
	\label{eq:cgtorque} 
\end{eqnarray}
where $U(R^N, \Omega^N)$ is the coarse-grained potential, $\bm{f}_i(\bm{r}^n) = -\frac{\partial\bm{u}}{\partial \bm{r}_i}$ is the force on fine-grained particle $i$, with $\bm{u}(\bm{r}^n)$ the fine-grained potential and $\langle \cdots\rangle_{R^N,\Omega^N}$ denoting an average over fined-grained configurations mapped to coarse-grained configuration $(R^N, \Omega^N)$. 

Consistency between the momentum-space distributions requires the mass $M_I$ of coarse-grained particle $I$ to be the sum of the masses of its constituent fine-grained particles, i.e.\cite{Nguyen2022}
\begin{equation}
	M_I = \sum_{i \in \zeta_I} m_i.
	\label{eq:cgmass}
\end{equation}
In addition, provided that the inertia tensor $\pmb{\mathbb{I}}_{\mathrm{FG},I}$ of the group of fine-grained particles mapped to this coarse-grained particle does not depend on the configuration of the other particles,\cite{Nguyen2022}
\begin{equation}
	I_{I,q}^{1/2}\exp \left(-\frac{I_{I,q}\omega_{I,q}^2}{2\kB T} \right) 
	\approx \left\langle 
	I_{\mathrm{FG},I,q}^{1/2} \exp \left(-\frac{I_{\mathrm{FG},I,q}\omega_{I,q}^2}{2\kB T } \right) \right\rangle_{\bm{R}_I,\bm{\Omega}_I},
	\label{eq:cginertia}
\end{equation}
where $I_{I,q}$, $I_{\mathrm{FG},I,q}$, and $\omega_{I,q}$ are the components of the coarse-grained moment of inertia, fine-grained moment of inertia, and angular velocity about the $q$ axis, and $\langle \cdots\rangle_{\bm{R}_I,\bm{\Omega}_I}$ denotes an equilibrium average of fine-grained configurations consistent with the coordinate mapping of coarse-grained particle $I$. Furthermore, if the fluctuations in $I_{\mathrm{FG},I,q}$ are small compared to its mean, it can be shown that\cite{Nguyen2022}
\begin{equation}
	I_{I,q} \approx \left\langle I_{\mathrm{FG},I,q} \right\rangle_{\bm{R}_I,\bm{\Omega}_I},
	\label{eq:cginertia2}
\end{equation}
i.e. the principal moment of inertia of a coarse-grained particle about each principal axis $q$ is approximately equal to the equilibrium average of the corresponding principal moment of the fine-grained particles mapped onto it. 

The AFM-CG method was derived only for the constant-volume conditions of the canonical ensemble, but is straightforwardly generalized to constant-pressure conditions by analogy with the MS-CG method for spherical coarse-grained particles in the isothermal-isobaric ensemble.\cite{das2010a} Thus, the force- and torque-matching conditions at constant pressure are the same as those in Eqs.~\eqref{eq:cgforce} and~\eqref{eq:cgtorque}, except that the coarse-grained forces, torques, and potential are in general functions of the coarse-grained system volume $V$ and the equilibrium average is constrained to configurations in which the fine-grained system volume $v = V$.  The coarse-grained potential must also satisfy a virial-matching condition,\cite{das2010a}
\begin{align} 
	W(\bm{R}^N,\bm{\Omega}^N,V) &= -\frac{\partial U}{\partial V} \nonumber\\
	&= \left\langle \frac{(n-N) \kB  T}{v} + \frac{1}{3v}\sum_{i = 1}^n {\bm{f}_i \cdot \bm{r}_i} \right\rangle_{R^N,\Omega^N,V} 
	\label{eq:cgvirial}
\end{align}

In summary, for the equilibrium phase-space distribution of the coarse-grained model to match that of the fine-grained model in the isothermal-isobaric ensemble, the coarse-grained potential should satisfy Eqs.~\eqref{eq:cgforce}, \eqref{eq:cgtorque}, and~\eqref{eq:cgvirial}, while the coarse-grained masses and principal moments of inertia should satisfy Eqs.~\eqref{eq:cgmass} and~\eqref{eq:cginertia}, respectively. As shown below, using the more approximate Eq.~\eqref{eq:cginertia2} to parameterize the moments of inertia gives almost the same results as Eq.~\eqref{eq:cginertia}, even for a flexible molecule, so we have used this simpler equation for parameterization later on.

\section{Methods}
\label{sec:methods}

\subsection{Force-, torque-, and virial-matching algorithm}

The analytical expression for the coarse-grain potential $U$ is not usually known. However, an approximation to the functional form can be obtained using a neural-network optimization algorithm with Eqs.~\eqref{eq:cgforce}, \eqref{eq:cgtorque}, and~\eqref{eq:cgvirial} acting as necessary constraints. In general, $U(\bm{R}^N,\bm{\Omega}^N,V)$ is a function of the particle configuration and system volume. In this work, we have assumed that $U$ does not depend explicitly on $V$, in which case\cite{das2010a}
\begin{equation}
	\frac{\partial U}{\partial V} =\frac{1}{3V}\sum_{I=1}^N \frac{\partial U}{\partial \bm{R}_I} \cdot \bm{R}_I .
	\label{eq:cgvirial2}
\end{equation} 
With this approximation, the virial-matching condition in Eq.~\eqref{eq:cgvirial} can be written, using $v=V$, as 
\begin{equation} 
	-\sum_{I=1}^N \frac{\partial U}{\partial \bm{R}_I} \cdot \bm{R}_I 
	= \left\langle 3(n-N) \kB  T + \sum_{i = 1}^n {\bm{f}_i \cdot \bm{r}_i} \right\rangle_{R^N,\Omega^N,V}. 
	\label{eq:cgvirial3}
\end{equation}
Despite this approximation, we show later on that the coarse-grained models parameterized accurately match the average density of the corresponding all-atom fine-grained system at constant pressure.

To ensure that all equivalent configurations are assigned the same position in coordinate space, a transformation was made from the set of Cartesian coordinates to a vector $\bm{D}_{IJ}$ that was invariant under translation, rotation, and permutation of any pair of coarse-grained particles $I$ and $J$, \cite{Bartok2013,Wang2018,Han2018,Zhou2019} which was defined in terms of the positions, $\bm{R}_I$  and $\bm{R}_J$, and orientations,   $\bm{\Omega}_I$ and $\bm{\Omega}_J$, of the two particles by
\begin{align}
	\bm{D}_{IJ} 
	&= \left\{ R_{IJ}, \bm{R}_{IJ} \cdot \bm{\Omega}_{I,1}, 
	\bm{R}_{IJ} \cdot \bm{\Omega}_{I,2}, \bm{R}_{IJ} \cdot \bm{\Omega}_{I,3},\right. \nonumber\\
	&\qquad\bm{R}_{IJ} \cdot \bm{\Omega}_{J,1}, \bm{R}_{IJ} \cdot \bm{\Omega}_{J,2}, 
	\bm{R}_{IJ} \cdot \bm{\Omega}_{J,3},  \nonumber\\
	&\qquad\bm{\Omega}_{I,1} \cdot \bm{\Omega}_{J,1}, 
	\bm{\Omega}_{I,1} \cdot \bm{\Omega}_{J,2}, \bm{\Omega}_{I,1} \cdot \bm{\Omega}_{J,3}, \nonumber\\ 
	&\qquad\bm{\Omega}_{I,2} \cdot \bm{\Omega}_{J,1}, \bm{\Omega}_{I,2} \cdot \bm{\Omega}_{J,2}, 
	\bm{\Omega}_{I,2} \cdot \bm{\Omega}_{J,3}, \nonumber\\
	&\qquad\left.\bm{\Omega}_{I,3} \cdot \bm{\Omega}_{J,1}, 
	\bm{\Omega}_{I,3} \cdot \bm{\Omega}_{J,2}, \bm{\Omega}_{I,3} \cdot \bm{\Omega}_{J,3}  \right\} ,
	\label{eq:Dvector}
\end{align}
where $R_{IJ} \equiv \lVert \bm{R}_{IJ}\rVert$, $\bm{R}_{IJ} \equiv \bm{R}_I - \bm{R}_J$ and $\bm{\Omega}_I$ and $\bm{\Omega}_J$ are specified by rotation matrices of the form of Eq.~\eqref{eq:cgorientation}. The coordinates of each neighbor within the cut-off distance of particle $I$ were transformed to a $\bm{D}_{IJ} $ vector. All the $\bm{D}_{IJ} $  vectors for a given neighborhood were concatenated into a 2D matrix $\pmb{\mathbb{D}}_I$ of size $N \times \dim(\bm{D}_{IJ})$ representing a unique configurational fingerprint for coarse-grained particle $I$. 

The potential function could then be written in terms of a set of neural network trainable parameters and activation functions transforming $\pmb{\mathbb{D}}_I$ to a potential energy value.  While $\pmb{\mathbb{D}}_I$ is a sufficient specification of the coarse-grained coordinates to enforce relevant invariant properties of the molecular environment, it does not possess all the symmetries of the potential energy surface that it aims to fit. \cite{Bartok2013,Gastegger2018} For each molecular environment, it was assumed that the interactions were predominantly short-ranged such that neighbors beyond a certain cut-off distance, $\Rc$, did not contribute to the potential.\cite{Behler2011} This condition can be enforced by a  cut-off function of the form 
\begin{eqnarray}
	\gc(R_{IJ}) =
	\begin{cases}
		\frac{1}{2} \bigg[ \cos \left( \frac{ \pi R_{IJ}}{\Rc} \bigg) +1 \right],      & \quad  R_{IJ} \leq \Rc, \\
		0,  & \quad  R_{IJ} > \Rc .
	\end{cases}
\end{eqnarray}
A set of these cut-off functions can enforce the radial symmetry conditions of the underlying potential energy surface by storing information about the radial distribution of neighbors according to\cite{Behler2011} 
\begin{eqnarray}
	G_{I}^1 = \displaystyle\sum_{J\neq I} { \gc (R_{IJ} ) } .
	\label{eq:G1}
\end{eqnarray}
Continuity of the potential along angular dimensions was ensured by using a compression layer to learn a set of collective variables from vector $\bm{D}_{IJ} $ which were constrained by the well-behaved modified $G^5$ symmetry function \cite{Behler2011} given by
\begin{eqnarray}
	G_{I}^5 = \displaystyle\sum_{J\neq I}\prod_{\mu=1}^{M} 2^{1-\nu}  \left(1 + \lambda \cos \theta_{IJ,\mu} \right)^{\nu}  e^{- \eta (R_{IJ}-\Rs)^2 }  \gc(R_{IJ}),
	\label{eq:G5}
\end{eqnarray}
where $\lambda \in \{-1,1\}$ and  $\Rs$, $\nu$, and $\eta$ are tunable hyperparameters, while $\{\cos \theta_{IJ,\mu}\}$, is the set of machine-learned collective variables with the same properties as the angular component of the underlying potential and $M$ is the total number of machine-learned angular variables. These angular symmetry functions store information about the angular-radial distribution of neighbors in the local environment of coarse-grained particle $I$. Unlike the case of spherically symmetric particles, in a local reference frame, a neighboring anisotropic particle requires a minimum of seven independent scalar variables to fully describe its position and orientation. However, previous implementations of analytical potentials, including the Gay--Berne potential, \cite{Gay1981,Berardi1995} have used fewer coordinates for the calculation of the potential and forces. Similarly, for the neural-network potential, an additional compression layer was included to remove the redundant angles from the  $\bm{D}_{IJ}$ vectors, since the combination of translation and rotation in 3D is parameterized by at most 7 unique coordinates.  The Behler symmetry functions were enforced on the output of the compression layer, ensuring that the learned compression had the same symmetry and continuity of the underlying potential. The reduction in the dimension of $\bm{D}_{IJ} $ also decreases the amount of data that is needed to train a sufficiently accurate potential. By removing the redundant angles in $\bm{D}_{IJ} $  there is a  reduced possibility of over-fitting on a small data set. 

A set of these symmetry functions with tuned hyperparameters $(\lambda,\nu,\eta,\Rs, \Rc)$ can be used to uniquely represent the structural fingerprint of the molecular environment. Symmetry functions used to represent the local environment were constructed using all possible permutations of values from a specified set of hyperparameters. Training of the neural network started with 8 symmetry functions and hyperparameters tuned to minimize the loss function, which is defined below. New symmetry functions were added to the set if they resulted in a significant reduction in the neural-network loss compared with the preceding iteration. The set of hyperparameters in the symmetry functions used in the anisotropic coarse-grained models parameterized in this work can be found in the Supplementary Material.

To further reduce the amount of data needed to train the neural network, a prior repulsive potential was defined with pairwise additive properties. This potential was used to ensure physical behavior in regions of the potential where the forces are large and thus are rarely sampled in an equilibrium MD simulation. This prior potential only needs to satisfy two conditions: firstly, it must be repulsive at short radial separations, and, secondly, the position of its repulsive barrier must be orientationally dependent. A simple equation satisfying these conditions is 
\begin{equation}
	U_{\mathrm{prior},I} = \sum_{J\neq I} B_1 \sigmac\left(\pmb{\mathbb{D}}_I\right)^{-B_2},
	\label{eq:Uprior}
\end{equation}
where $\sigmac$ is a neural-network compression layer function and $B_1$ and $B_2$ are strictly positive trainable parameters. It is also possible to achieve a similar large repulsive barrier through a more advanced nonlinear sampling of the MD simulation data. The prior potential fits the purely repulsive part of the angular-dependent potential to the molecular environment, while the neural-network potential fits the attractive and oscillatory corrections to the environment. The final prediction for the potential energy of the environment of coarse-grained particle $I$ is therefore the sum of the neural-network potential $U_{\mathrm{NN},I}$ and the prior repulsive potential $U_{\mathrm{prior},I}$, \cite{Wang2019} 
\begin{align}
	U_I = U_{\mathrm{NN},I} + U_{\mathrm{prior},I},
\end{align}
and, thus, the total coarse-grained potential is
\begin{align}
	U=\sum_{I=1}^N U_I
\end{align}

From the matching conditions in Eqs.~\eqref{eq:cgforce}, \eqref{eq:cgtorque}, and \eqref{eq:cgvirial3}, optimization of the neural-network weights and biases requires a loss function of the form
\begin{align} 
	L &= \left\langle \sum_{I=1}^N\left(  \alpha\left\lvert \FFGI + \frac{\partial U}{\partial R_I}   \right\rvert^2 
	+ \beta  \left\lvert \tauFGI + \sum_{q} \Omega_{I,q}\times \frac{\partial U}{\partial \Omega_{I,q}} \right\rvert^2 \right)\right.  \nonumber\\
	&+ \left. \gamma  \left\lvert 3\left(n-N\right) \kB  T +\sum_{I=1}^N\left( \WFGI +  \frac{\partial U}{\partial \bm{R}_I} \cdot \bm{R}_I \right)\right\rvert^2 \right\rangle_{R^N,\Omega^N,V}  ,
	\label{eq:cgloss}
\end{align}
where 
\begin{equation}
	\FFGI \equiv \sum_{i \in \zeta_I } {\bm{f}_i},\; 
	\tauFGI \equiv \sum_{i \in \zeta_I } {\Delta \bm{r}_i \cross\bm{f}_i},\; 
	\WFGI \equiv  \sum_{i \in \zeta_I } {\bm{f}_i \cdot \bm{r}_i},  \label{eq:fgforce}
\end{equation}
and $\alpha, \beta $, and $\gamma $ are weights. These weights specify the fraction of each loss that is used for backpropagation and were free to change with the learning rate during optimization.\cite{Zhang2018}  Even though there have been significant efforts to develop methods to fit the constrained averaged coarse-grained forces directly, \cite{Ciccotti2005,Abrams2008} the average total fine-grained forces subject to the constraint of matching fine-grained and coarse-grained configurations are not easily obtained.
An indirect means of minimizing the loss function in Eq.~\eqref{eq:cgloss} above is possible by replacing the constrained ensemble average with an average over instantaneous unconstrained simulation configurations,\cite{Zhang2018}
\begin{align} 
	\Linst &= \sum_{t=1}^{N_\mathrm{t}}\left[\sum_{I=1}^N\left(\alpha\left\lvert \FFGI(\bm{r}_t^n) + \frac{\partial U(\bm{\xi}_t)}{\partial R_I}   \right\rvert^2 \right.\right. \nonumber \\
 &+ \left.\beta  \left\lvert \tauFGI(\bm{r}_t^n) + \sum_{q} \Omega_{I,q}(\bm{\xi}_t))\times \frac{\partial U(\bm{\xi}_t))}{\partial \Omega_{I,q}} \right\rvert^2 \right)  \nonumber\\
&+ \left. \gamma  \left\lvert 3\left(n-N\right) \kB  T
+ \sum_{I=1}^N\left( \WFGI(\bm{r}_t^n) +  \frac{\partial U(\bm{\xi}_t)}{\partial \bm{R}_I} \cdot \bm{R}_I(\bm{\xi}_t) \right)\right\rvert^2 \right],
\label{eq:cgloss2}
\end{align}
since it can be shown, for a sufficiently large dataset that comprehensively samples the equilibrium ensemble of the fine-grained system, that $L$ and $\Linst$ have the same global minimum. Here, $N_\mathrm{t}$ is the number of simulation configurations in the dataset, $\bm{r}_t^n$ and $v_t$ are the fine-grained coordinates and system volume for configuration $t$, and $\bm{\xi}_t = (\bm{R}^N(\bm{r}_t^n), \bm{\Omega}^N(\bm{r}_t^n), V(v_t))$ is the mapped coarse-grained configuration for this fine-grained configuration. The loss function was optimized using the minibatch gradient descent as implemented in TensorFlow. 

\begin{figure}
	\centering
	\includegraphics{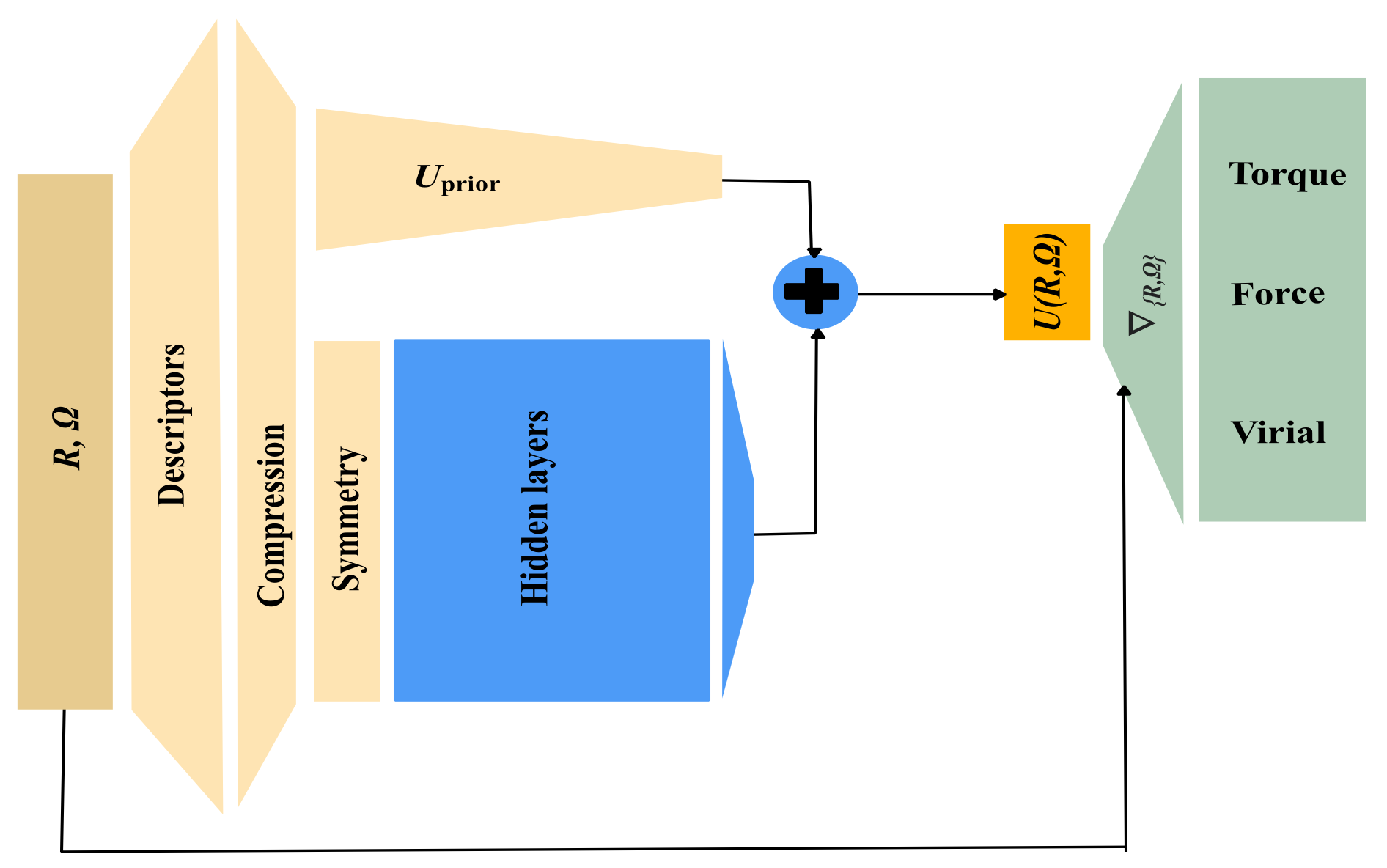}
	\caption{Schematic of anisotropic force-matching neural network architecture.}
	\label{fig:neuralnetwork}
\end{figure}

The feedforward neural network shown in Fig.~\ref{fig:neuralnetwork} was then trained, where the forward propagation used matrix $\pmb{\mathbb{D}}_I$ as an input to predict the coarse-grained potential $U$, after which TensorFlow's computational derivative was used to calculate the outputs, namely the predicted forces, torques, and virial. In the backpropagation stage, the loss function was used to calculate the error between the true and predicted values, which was then used to update the network weights and biases. The errors between the true and predicted parameters were calculated using TensorFlow's mean squared error, and gradient descent was implemented using TensorFlow's Adam optimizer. \cite{Kingma2014} Once the error of the neural network was minimized, the neural-network model was used to predict the forces, torques, and virial. However, removing the output and derivative layers gives access to the predicted potential of mean force. By optimizing the partial derivatives of the potential instead of the potential itself, by the nature of the derivative, there will be less oscillation in the potential at the edges of the data set close to the cut-off distances.

\subsection{LAMMPS modification and neural network implementation}

The neural network was constructed in Tensorflow (version 2.3.0) \cite{Abadi2016} using the Keras (version 2.4.3) functional API \cite{Chollet2015} and saved using the Tensorflow SavedModel format. The trained neural network was implemented in LAMMPS using the Tensorflow C API and cppflow wrapper. All simulations were carried out using the LAMMPS MD software package (version 20Nov19). \cite{Plimpton1995,Brown2011,Brown2012} The Optimized Potentials for Liquid Simulations-All Atom (OPLS-AA) force field \cite{Jorgensen1996,Jorgensen1998,rizzo1999,Price2001} was used for all all-atom simulations with a cut-off distance of 10~\AA\ for short-ranged non-bonded interactions; long-ranged electrostatic interactions were calculated with the particle--particle particle--mesh (PPPM) method \cite{Brown2012,Hockney1998} The bonds that include hydrogen were constrained using the SHAKE algorithm. \cite{ryckaert1977numerical} Simulations were carried out in the isothermal-isobaric (NPT) ensemble at a pressure of 1~atm, with the temperature and pressure controlled by a Nos\'e-Hoover thermostat and barostat. \cite{hoover1985canonical,nose1984molecular}

Neural-network training was carried out using data from a 25~ns all-atom simulation in which simulation configurations and forces and velocities were saved at 2~ps intervals. The simulation snapshots from the last 20~ns were shuffled and then divided into 4 groups of equal size, $\{g_0, g_1, g_2, g_3\}$. The neural network was initially trained on $g_0$ and validated on $g_3$. The validation set $g_3$ was further divided into an 8:2 ratio where the lesser was reserved as the test set. New snapshots were added from $g_1$ and $g_2$ if the mean errors of their predicted forces and torques were larger than that of the test set.  The accuracy of the trained neural network was then compared to the expected accuracy determined from k-fold cross-validation.\cite{Marcot2021,Bengio2003} During k-fold cross-validation, the last 20~ns of simulation data was shuffled and divided into 10 folds, $\{\psi_0, ..., \psi_9\}$. The model was validated on $\psi_i$ and trained on $\bigcup_{j \neq i}\psi_j$ for all $i,j \in \{0,1,\ldots,9\}$. The loss of the iterative training method was found to be identical to the k-fold cross-validation loss.

The coarse-grained simulations were done using a modified version of LAMMPS in which the trained neural network was introduced to calculate the forces and energies. The dimensions of the coarse-grained sites used in the simulations were derived from the inertia tensor of the all-atom model. To test the ability of the coarse-grained model to capture the properties of the all-atom model under a variety of conditions in addition to the single temperature at which the neural network was trained, the equilibrium structural properties of equivalent coarse-grained and all-atom systems were compared in simulations at several different temperatures. In all cases, the total length of the coarse-grained simulation was 25 ns, with the last 20 ns used to calculate structural and thermodynamic properties. 

\section{Results and Discussion}

\begin{figure}
\centering
\includegraphics{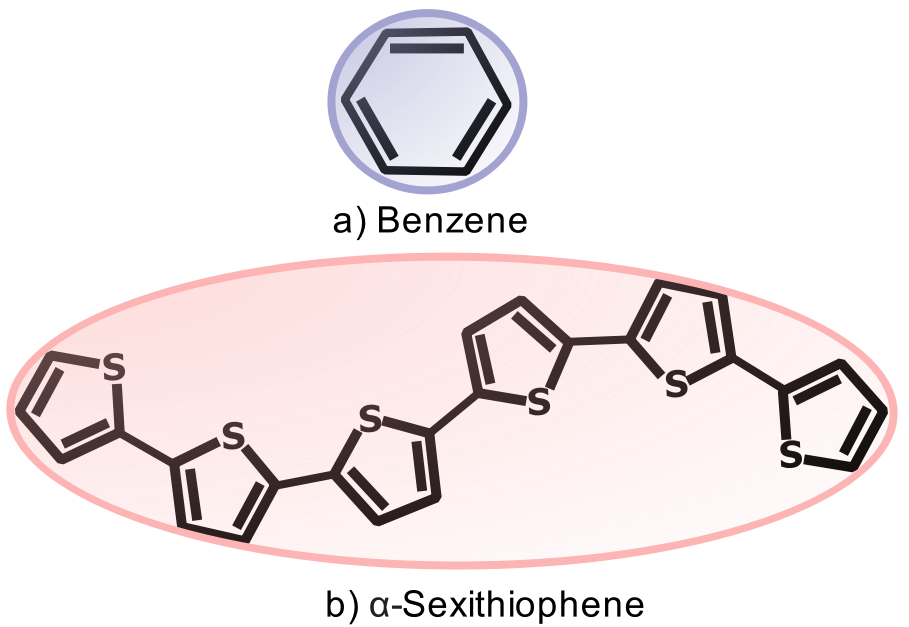}
\caption{Chemical structures of (a) benzene and (b) $\alpha$-sexithiophene with a coarse-grained ellipsoid superimposed on a possible configuration of each molecule.}
\label{fig:benzene_sexithiophene_structures}
\end{figure}

To demonstrate the flexibility of the method, we have used our neural-network model to construct coarse-grained interaction potentials for benzene, an archetypal anisotropic small molecule, and $\alpha$-sexithiophene, an organic semiconductor with significant applications in organic electronic devices\cite{katz1997organic,Fichou2000,dong2020orientation} (Fig.~\ref{fig:benzene_sexithiophene_structures}). These molecules were selected to demonstrate the neural network's ability to handle anisotropic molecules of varying complexity, flexibility, and aspect ratio while still reproducing the structural and phase behavior. 

The shape of a coarse-grained particle obtained from the anisotropic coarse-graining method is determined by the "average" shape of the fine-grained molecule or molecular fragment that is mapped to it under the parameterization conditions. Thus, the variation of the aspect ratio of the molecule or molecular fragment with temperature in the all-atom simulations can potentially be a qualitative indicator of the temperature transferability of the coarse-grained model. Here, the aspect ratio of the molecule was calculated as the ratio of the length to the breadth of the molecule, where the length was defined as the longest principal axis and the breadth was defined as the sum of the remaining two semi-axes. Unlike benzene, the thiophene--thiophene dihedral angles also have a temperature-dependent effect on the aspect ratio of sexithiophene. 

Neural networks in general are very good at interpolation but struggle with extrapolation. \cite{Haley1992,Na2022,Ding2021,Rigol2001} The accuracy of the model is therefore expected to decrease as the aspect ratio of the molecule deviates from that at the parameterization temperature, as well as when the density distribution is sufficiently different from the parameterization temperature. By parameterizing the systems in the liquid phase, the model can capture a wider variety of fluctuations in the density of the system and the dimensions of the molecules. The average size of a flexible molecule in the isotropic phase will be different from the size of the molecule when locked in a rigid crystal structure. \cite{Mueller2015,Seaton2006} However, this temperature-dependent size difference should decrease with increased rigidity of the molecule.

\subsection{Benzene}

Simulations consisting of 500 benzene molecules were carried out at 280, 300, 320, 330, and 350~K, and the coarse-grained neural-network model was parameterized at 300~K. The time step was 2~fs in the all-atom simulations and 12~fs in the coarse-grained simulations. The cut-off distance hyperparameter $\Rc$ was 10~\AA. The root mean squared validation error for the forces was \SI{2.55}{\kcal\per\mole\per\angstrom} and that of the torque was \SI{4.35}{\kcal\per\mol}. The average post-training error in the pressure was 0.0092~atm. 

\begin{figure}
	\centering
	\includegraphics{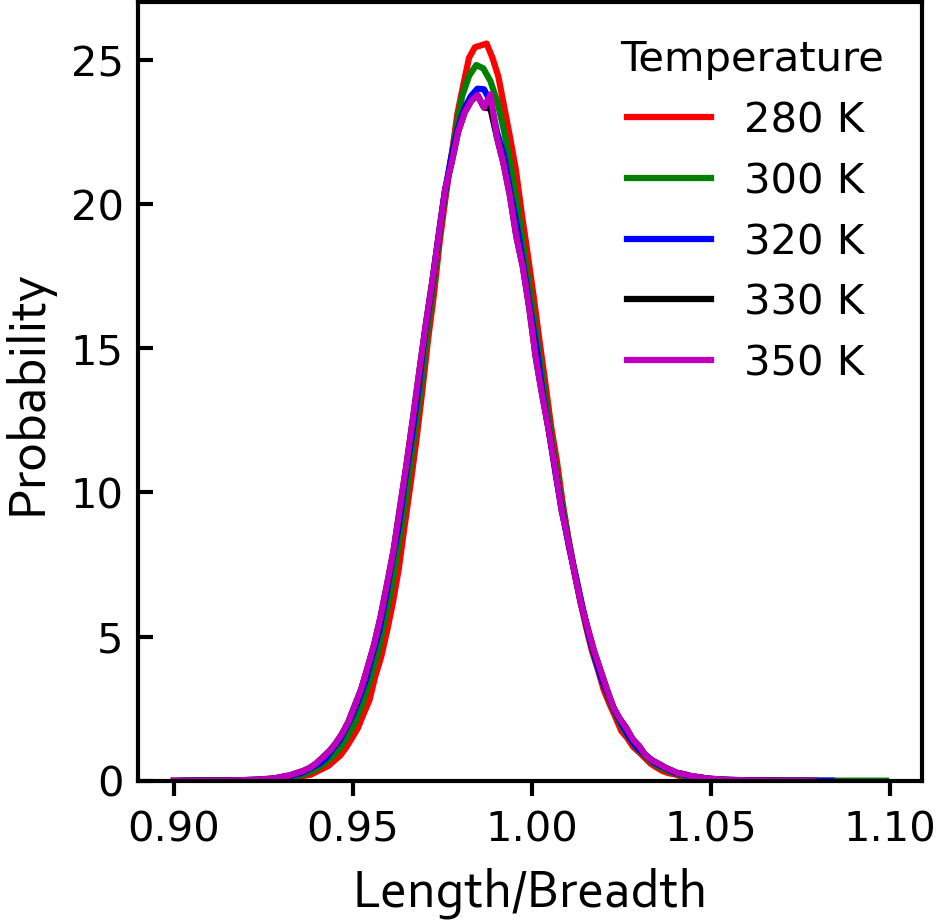}
	\caption{Length-to-breadth ratio of the all-atom benzene model at 1~atm and various temperatures.}
	\label{fig:benzeneAspectRatio}
\end{figure}

The average principal moments of inertia in the all-atom simulation at 300~K were used to determine the principal moments of the coarse-grained model using Eqn.~(\ref{eq:cginertia2}) (values given in the Supplementary Material) since fluctuations in the moments at the parameterization temperature were small. \cite{Nguyen2022} The variation of the molecular aspect ratio of the all-atom benzene model with temperature is shown in Fig.~\ref{fig:benzeneAspectRatio}. The distribution of possible dimensions observed for benzene is narrow and remains fairly constant with temperature, making benzene an ideal case where molecular flexibility does not contribute significantly to the overall error of the model.\cite{Sinitskiy2012}

\begin{figure}
	\centering
	\includegraphics{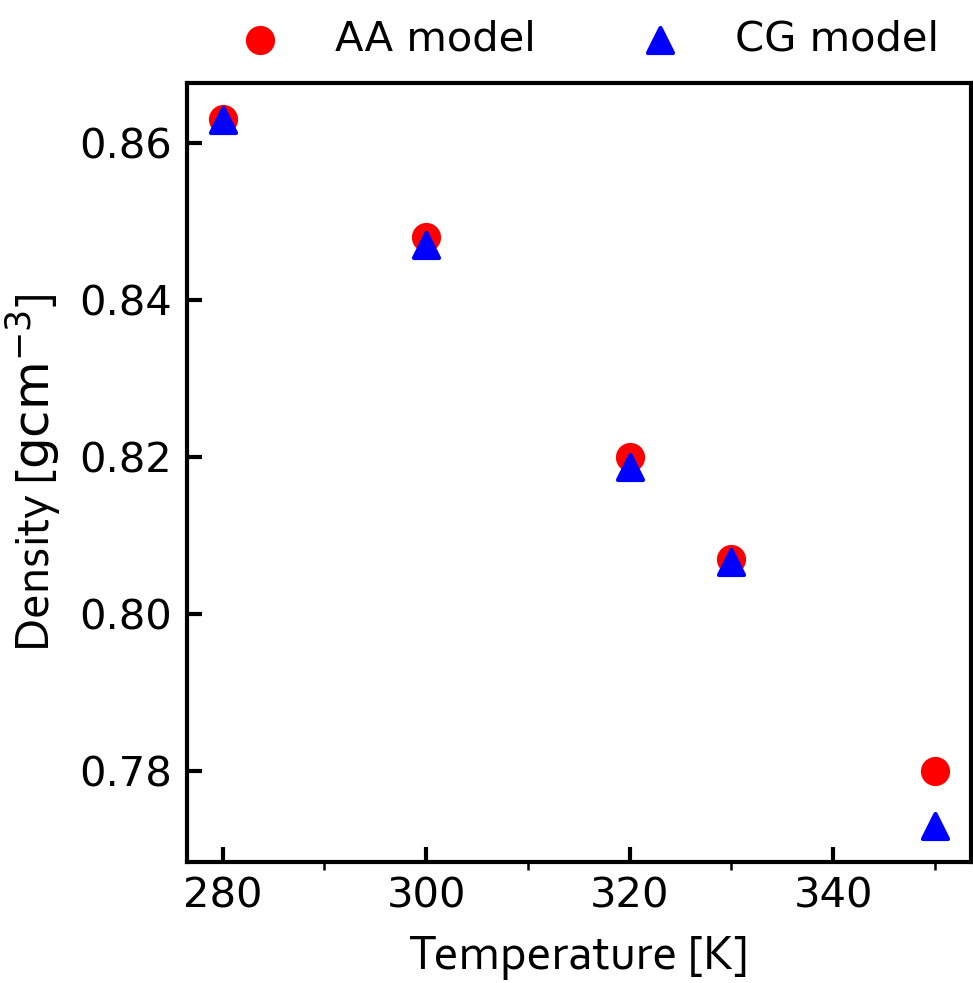}
	\caption{Density versus temperature of the all-atom (AA) and coarse-grained (CG) benzene models at 1~atm. Error bars are smaller than the symbols.}
	\label{fig:benzeneDensity}
\end{figure}

\begin{figure}
	\centering
	\includegraphics{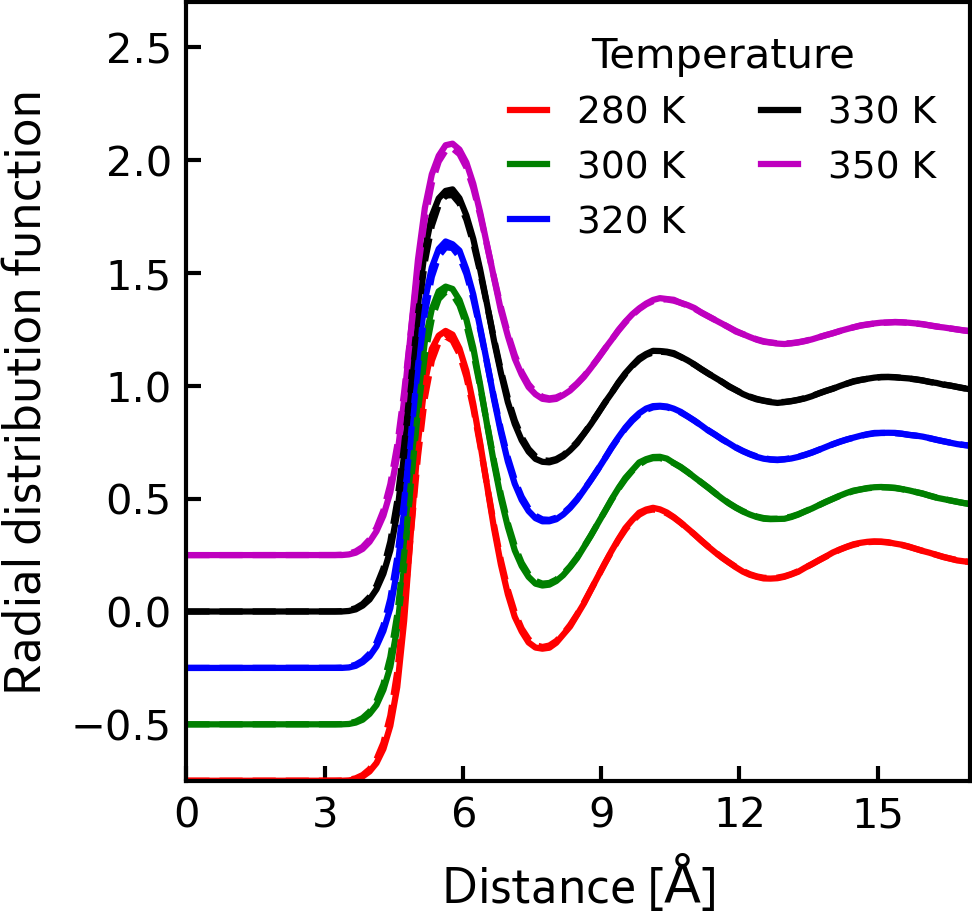}
	\caption{Radial distribution function (RDF) of the all-atom (solid lines) and coarse-grained (dashed lines) benzene models at 1~atm and various temperatures. The RDFs have been shifted vertically for clarity. }
	\label{fig:benzenerdf}
\end{figure}

Fig.~\ref{fig:benzeneDensity} shows that the coarse-grained neural-network model accurately captures the liquid density of the all-atom model over a wide range of temperatures from just above the freezing point to just below the boiling point, with only slight deviations for the temperature furthest from the parameterization temperature. As shown in Fig.~\ref{fig:benzenerdf}, the coarse-grained model also accurately predicts the radial distribution function (RDF) of the all-atom model over the same temperature range.

\begin{figure*}
	\centering
	\includegraphics{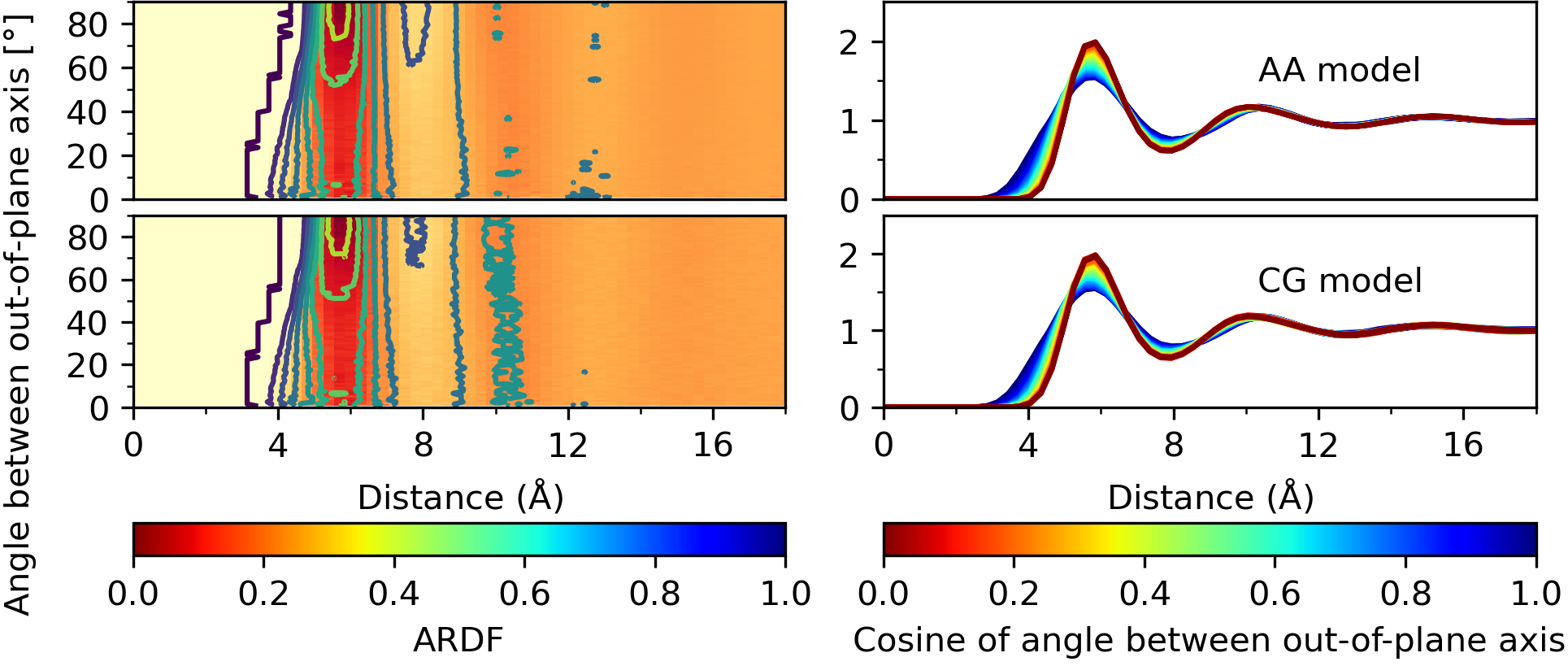}
	\caption{Angular--radial distribution function (ARDF) of the all-atom (AA) (top) and coarse-grained (CG) (bottom) benzene models at 300~K and 1~atm depicted as a heat map (left) and 1D slices at constant angle (right). Face-to-face, edge-to-edge, and parallel displaced configurations occur when the angle is 0\textdegree, while edge-to-face configurations occur at 90\textdegree.}
	\label{fig:benzeneardf}
\end{figure*}

To further elucidate the accuracy of the neural-network coarse-grained model, the angular--radial distribution function (ARDF) was analyzed. The ARDF is defined by 
\begin{equation}
	g(r,\theta) = \frac{\langle n(r,\theta) \rangle}{\frac{4}{3} \pi \rho [(r + \Delta r)^3 - r^3] \sin \theta \Delta \theta} ,
\end{equation}
where $\langle n(r,\theta) \rangle$ is the average number of molecules in the spherical shell within the bounds $r$ to  $r + \Delta r$ of the center-of-mass of a chosen molecule and having an out-of-plane axis rotation of $\theta$ with respect to the out-of-plane axis of the chosen molecule, \cite{Falkowska2016} and $\rho$ is the bulk number density. Fig.~\ref{fig:benzeneardf} shows the 2D heatmap of the ARDF along with 1D slices of this function at specific angles at 300~K (the parameterization temperature) for the all-atom and coarse-grained models. The ARDFs at the other simulated temperatures are compared in the Supplementary Material. At all simulated temperatures between 280 and 350~K, the coarse-grained model captures all the major features of the fine-grain structure of the fluid. The only difference is a slight underestimation of the peak heights by the coarse-grained model. The neural-network model is, however, able to more faithfully capture the angular radial distribution of benzene at all temperatures compared with the coarse-grained benzene model previously parameterized with the AFM-CG method using a pair potential to describe the interparticle interactions. \cite{Nguyen2022} This improvement can be attributed to the greater flexibility of the neural-network potential in describing the intermolecular interactions. The neural-network model can demonstrate temperature transferability through careful selection of the neural network hyperparameters to prevent overfitting of the local number density variations. 

The coarse-grained simulation of anisotropic molecules using a neural-network potential is more suited for large, preferably rigid, molecules, for which a high degree of coarse-graining can be achieved with reasonable accuracy. However, the model was still able to achieve a modest $20 \times$ speedup compared with the atomistic simulations, through a combination of reduced computation time per timestep and a larger timestep. This poor performance for a small molecule such as benzene is due to the small reduction in the number of degrees of freedom from the all-atom model to the coarse-grained model, coupled with a neural-network potential that is more computationally expensive than an analytical potential. Nevertheless, computational savings are obtained even in this suboptimal case. Simulations were carried out on a 4-core Intel i7-4790K CPU, but, further speedups could be achieved by taking advantage of the GPU-enabled version of TensorFlow.

\subsection{Sexithiophene}

Simulations of 512 sexithiophene molecules were carried out at 570, 590, 640, and 680~K temperatures, corresponding to temperatures previously identified in all-atom MD simulations to correspond to crystalline (K), smectic-A (Sm-A), nematic (N), and isotropic (I) phases respectively. \cite{Pizzirusso2011}  The time step was 1~fs in the all-atom simulations and 12~fs in the CG simulations. Although we have used the OPLS-AA force field for our all-atom simulations, whereas these previous MD simulations \cite{Pizzirusso2011} used the related AMBER force field \cite{weiner1984new,weiner1986all,cornell1995second} the structural properties of systems simulated with these two force fields (in particular the density, orientational order parameter, and radial distribution function discussed below) are very similar for the temperature range studied.  The cut-off distance hyperparameter $\Rc$ was set to 21~\AA. The neural-network model was parameterized using simulation snapshots from the isotropic phase at 680~K, where the molecular mobility was highest. The conditions of the isotropic bulk phase are advantageous in efficiently sampling the configuration space, especially rare high-energy configurations necessary for the accurate reproduction of the repulsive part of the coarse-grained potential. 

\begin{figure}
	\centering
	\includegraphics{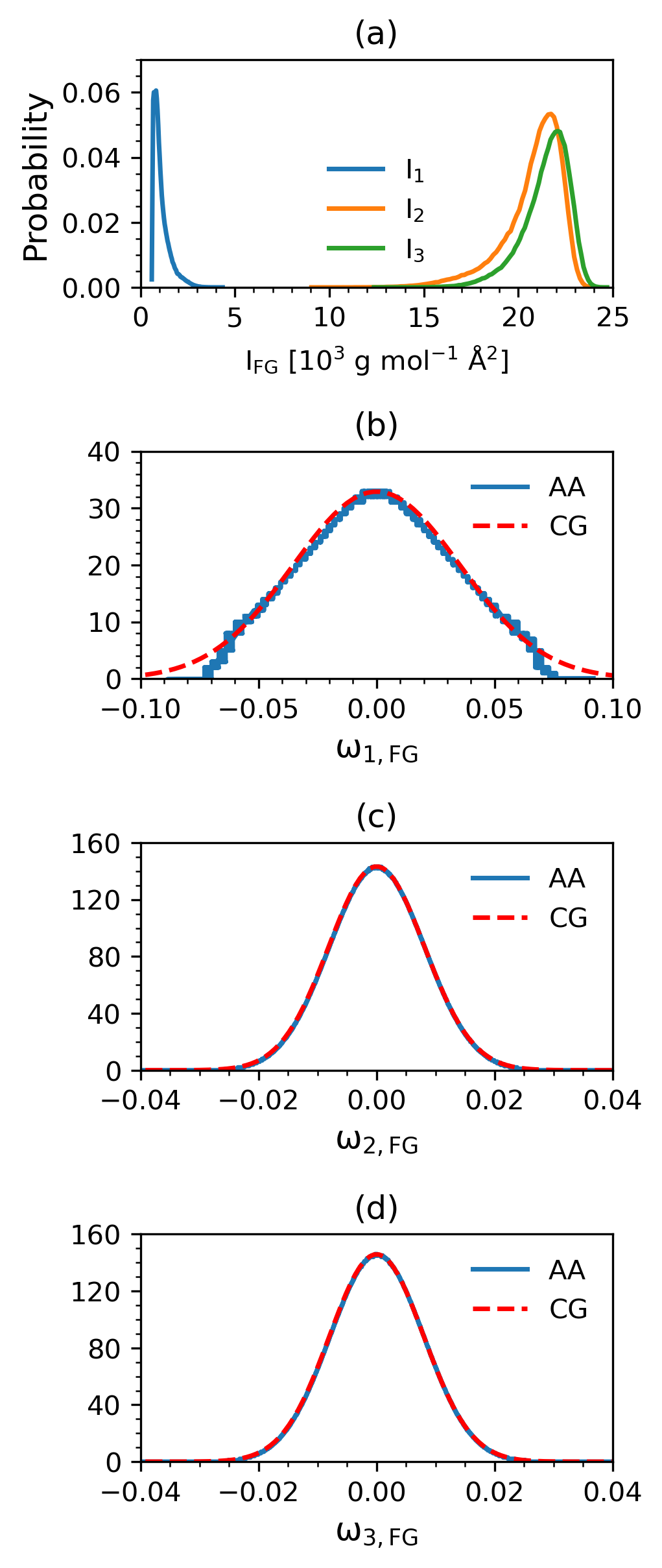}
	\caption{(a) Principal moment of inertia distributions for  all-atom (AA) sexithiophene model at 680~K and 1~atm. The corresponding angular velocity distributions of each principal axis along with the coarse-grained (CG) fit to the distribution given by Eq.~(\ref{eq:cginertia}) is shown in (b)--(d).}
	\label{fig:sexithiopheneinertiadistribution2}
\end{figure}

As shown in Fig.~\ref{fig:sexithiopheneinertiadistribution2}a, the distributions of the principal moments of inertia of sexithiophene in the all-atom simulation at the parameterization temperature are broad, indicating that Eqn.~\eqref{eq:cginertia2} may not be adequate for parameterizing the moments of inertia of the coarse-grained model. However, we found that using the more general Eqn.~\eqref{eq:cginertia} to parameterize the coarse-grained moments of inertia (by fitting the distributions in Fig.~\ref{fig:sexithiopheneinertiadistribution2}b--d gave values within <$1\%$. So we used the values from Eqn.~\eqref{eq:cginertia2} in the coarse-grained model.

The root mean squared validation error for the sexithiophene forces was \SI{3.95}{\kcal\per\mole\per\angstrom} and that of the torque was \SI{9.8}{\kcal\per\mole}. The sexithiophene final force and torque losses were larger than those of benzene because the model was not complex enough to account for the bending of the polymer and the rotation of the individual thiophene rings. The loss is also skewed to larger values when compared with benzene because sexithiophene is a larger molecule and so the interactions between molecules are stronger overall. 

\begin{figure}
	\centering
	\includegraphics{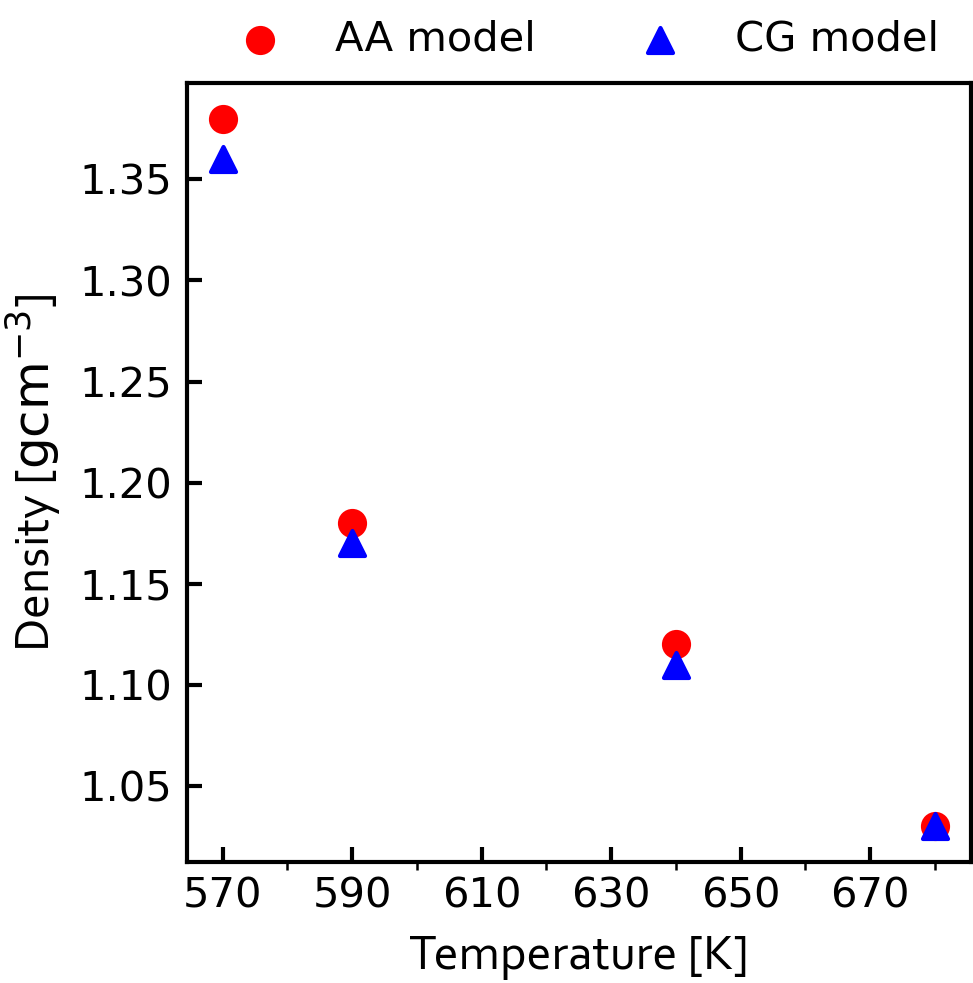}
	\caption{Density versus temperature of the all-atom (AA) and coarse-grained (CG) sexithiophene models at 1~atm.  Error bars are smaller than the symbols.}
	\label{fig:sexithiopheneDensity}
\end{figure}

\begin{figure}
	\centering
	\includegraphics{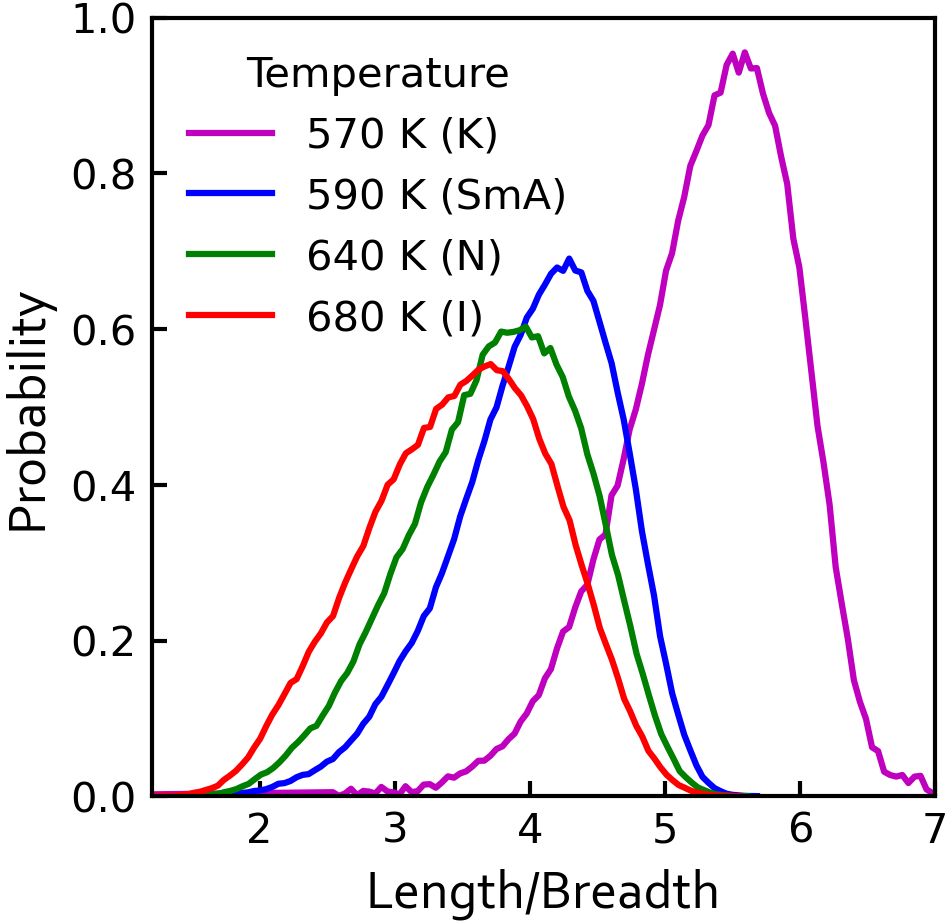}
	\caption{Length-to-breadth ratio of all-atom sexithiophene model at 1~atm  and various temperatures. The simulated phase is given in parentheses after each temperature in the legend (I = isotropic, N = nematic, SmA = smectic A, K = crystal).} 
	\label{fig:sexithiopheneAspectRatio}
\end{figure}

The structural properties of the coarse-grained model were compared with those of its all-atom counterpart at each of the simulated temperatures. The nonlinear change in density with respect to temperature is associated with the phase changes that occur at the simulated temperatures (Fig.~\ref{fig:sexithiopheneDensity}). \cite{Pizzirusso2011} The density of the coarse-grained system agrees well with that of the all-atom system, with minimal deviations from the fine-grained system with increasing distance from the parameterization temperature. Compared with benzene, sexithiophene has a much larger change in density between the crystalline and the isotropic phase. This difference results in less overlap between the local density variations in the crystalline phase at the lowest temperature and the training data set in the isotropic phase at the highest temperature. The sexithiophene molecule is also much more flexible than benzene, as seen in the wide distribution of the aspect ratio in the all-atom model at all the simulated temperatures shown in Fig.~\ref{fig:sexithiopheneAspectRatio},  and its dimensions change significantly with temperature over the range studied. Another limitation of representing sexithiophene as a single-site ellipsoid is the loss of thiophene--thiophene torsional information.  That is, for any given position and orientation of the coarse-grained ellipsoid there are multiple different relative orientations between the thiophene groups.  \cite{Tsourtou2018a} This loss of information is significant because the anisotropic interactions of the thiophene subunits are lost, which reduces the neural network's ability to isolate which of the two short axes corresponds to the $\pi$-stacking direction.

\begin{figure}
	\centering
	\includegraphics{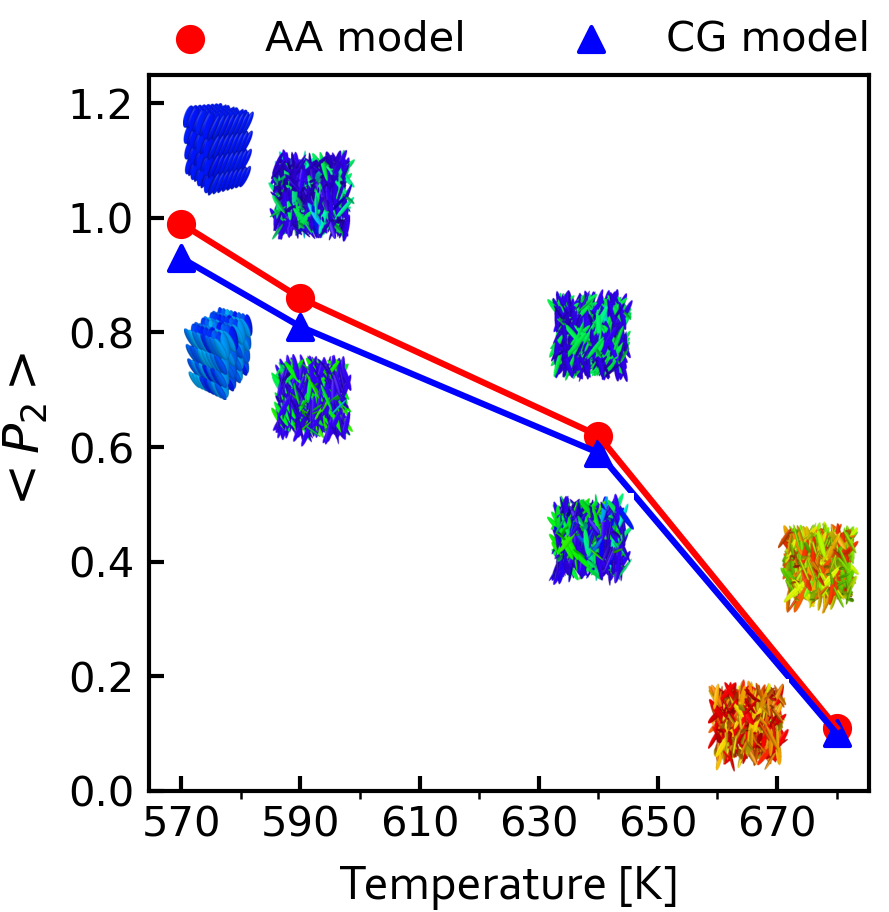}
	\caption{Orientational order parameter versus temperature for the all-atom (AA) and coarse-grained (CG) sexithiophene models at 1~atm. Typical simulation configurations are shown at each temperature for each system (AA model above the data points and CG model below), in which the molecules have been colored according to their orientation with respect to the phase director (blue = parallel, red = perpendicular). Error bars are smaller than the symbols.}
	\label{fig:sexithiopheneOrderParameter}
\end{figure}

To further confirm that the density changes were associated with transitions from the crystalline phase through the nematic and smectic phases to the isotropic phase, the scalar orientational order parameter $P_2$ was introduced. 
For a given simulation snapshot at time $t$, $P_2$ can be found by diagonalizing the ordering matrix
\begin{eqnarray}
	\bm{Q} = \frac{1}{2N}\sum_{I=1}^{N}{\left(3 \bm{u}_I \otimes \bm{u}_I -\bm{E} \right)} ,
\end{eqnarray}
where $\bm{u}_I$ is the unit vector along the molecular axis and  $\bm{E}$ is the identity matrix. $\langle P_2 \rangle$ is the average over the largest eigenvalue of this matrix for all snapshots of equilibrium configurations. \cite{Pizzirusso2011} Larger values of the scalar orientational order parameter close to one indicate an ordered crystalline structure while values close to zero correspond to an isotropic disordered phase. The coarse-grained model reproduces the orientational order parameter of the all-atom model reasonably well over the temperature range simulated, as shown in Fig.~\ref{fig:sexithiopheneOrderParameter}. The coarse-grained model underestimates the degree of orientational ordering observed in the all-atom model away from the parameterization temperature, likely because it does not capture the increasing molecular shape anisotropy that is observed in the all-atom model as the temperature decreases (Fig.~\ref{fig:sexithiopheneAspectRatio}). As expected, the largest difference occurs in the predicted crystalline phase. 

\begin{figure}
	\centering
	\includegraphics{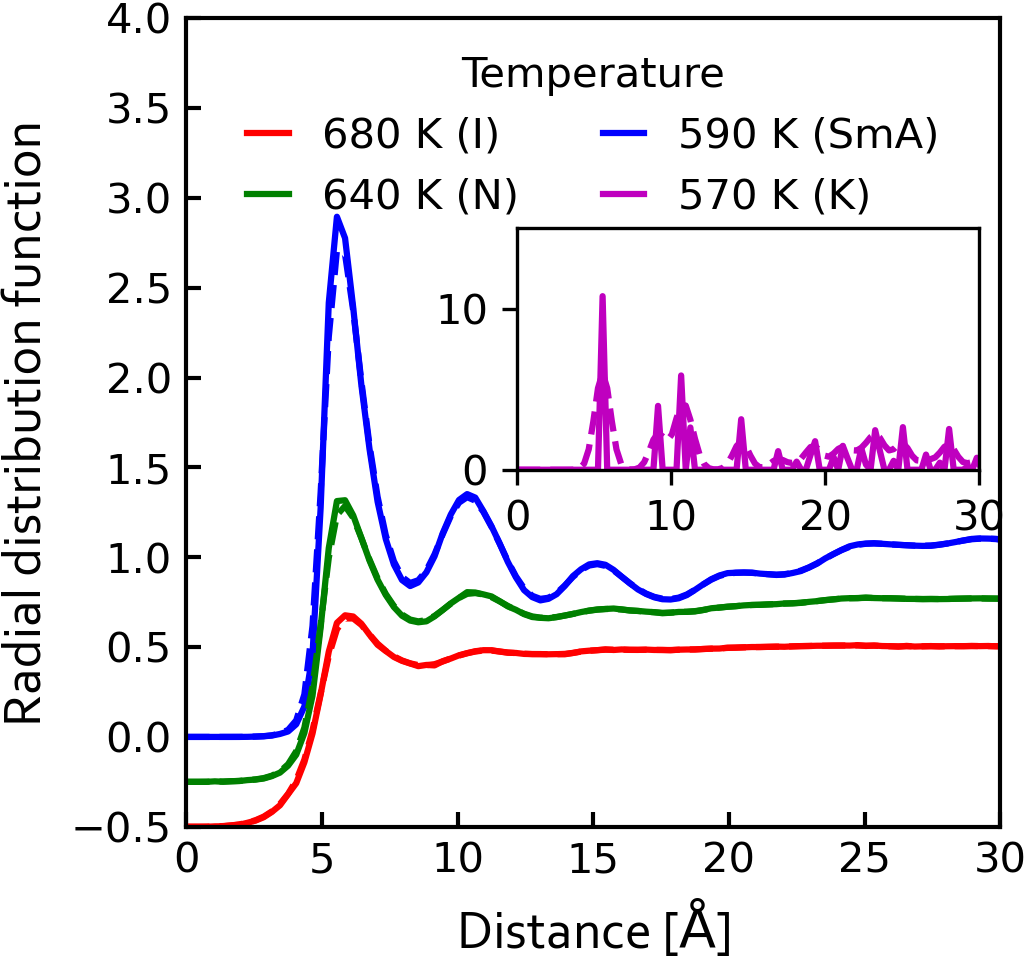}
	\caption{Radial distribution function (RDF) of the all-atom (solid lines) and coarse-grained (dashed lines) sexithiophene models at 1~atm and various temperatures. The RDFs have been shifted vertically for clarity. The simulated phase is given in parentheses after each temperature in the legend (I = isotropic, N = nematic, SmA = smectic A, K = crystal).}
	\label{fig:sexithiopheneRdf}
\end{figure}

The same trend is seen in the radial distribution functions shown in Fig.~\ref{fig:sexithiopheneRdf}, in which the agreement between the coarse-grained and all-atom models at most temperatures is excellent, with the largest deviations for the crystalline phase. The underestimation and broadening of the peaks in the crystalline radial distribution function explain the discrepancy between the order parameter of the all-atom and coarse-grained models. The observed differences are most likely due to the effect on molecular packing of the aforementioned discrepancy in molecular shape between the two models as temperature decreases. \cite{Xia2019} Nevertheless, even in the crystalline phase, the coarse-grained model captures the peak positions of the radial distribution function very well.

\begin{figure*}
	\centering
	\includegraphics{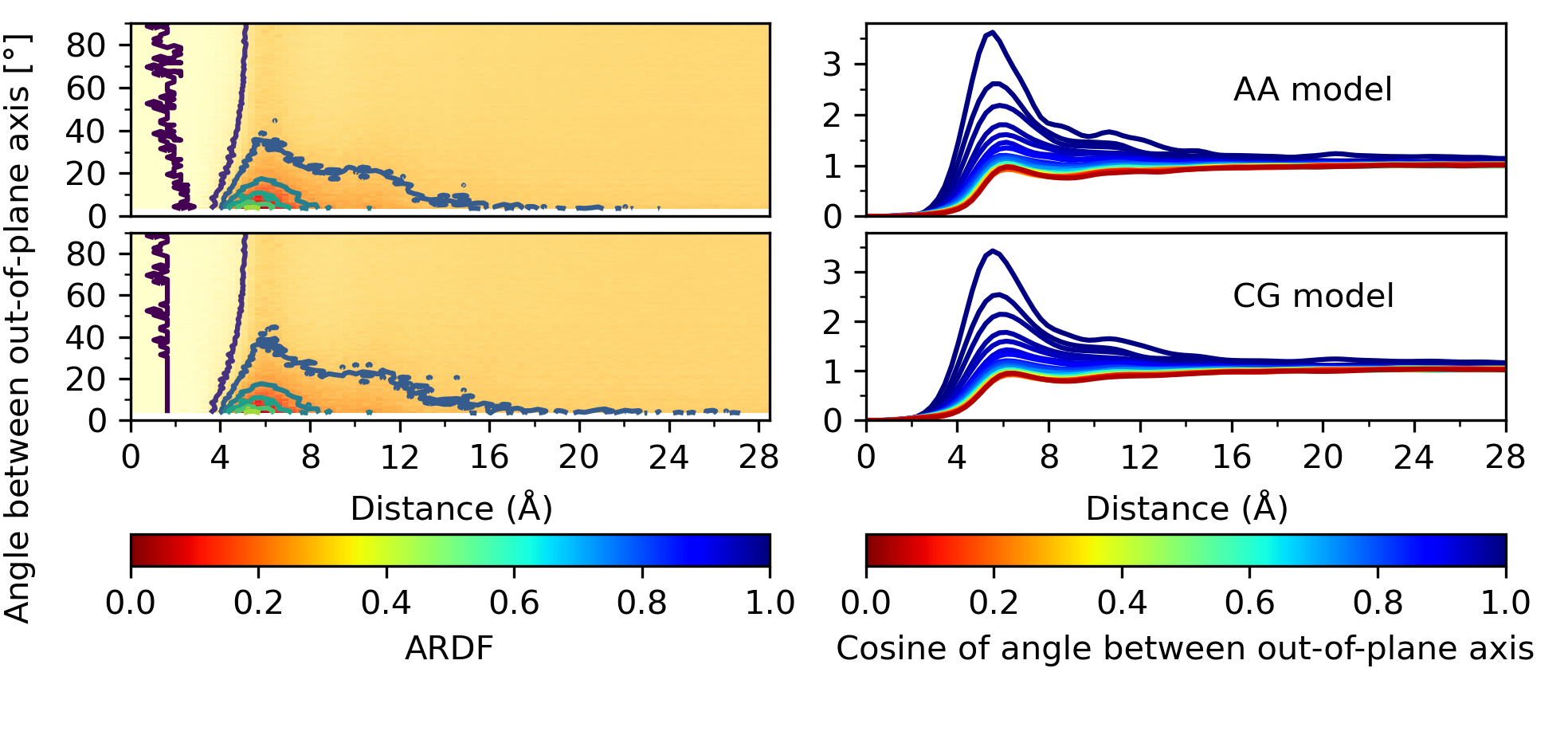}
	\caption{Angular-radial distribution function (ARDF) of the all-atom (AA) (top) and coarse-grained (CG) (bottom) sexithiophene models at 680~K and 1~atm (isotropic phase)  depicted as a heat map (left) and 1D slices at constant angle (right). Face-to-face, edge-to-edge, and parallel displaced configurations occur when the angle is 0\textdegree, while edge-to-face configurations occur at 90\textdegree.}
	\label{fig:sexithiopheneArdf}
\end{figure*}

\begin{figure*}
	\centering
	\includegraphics{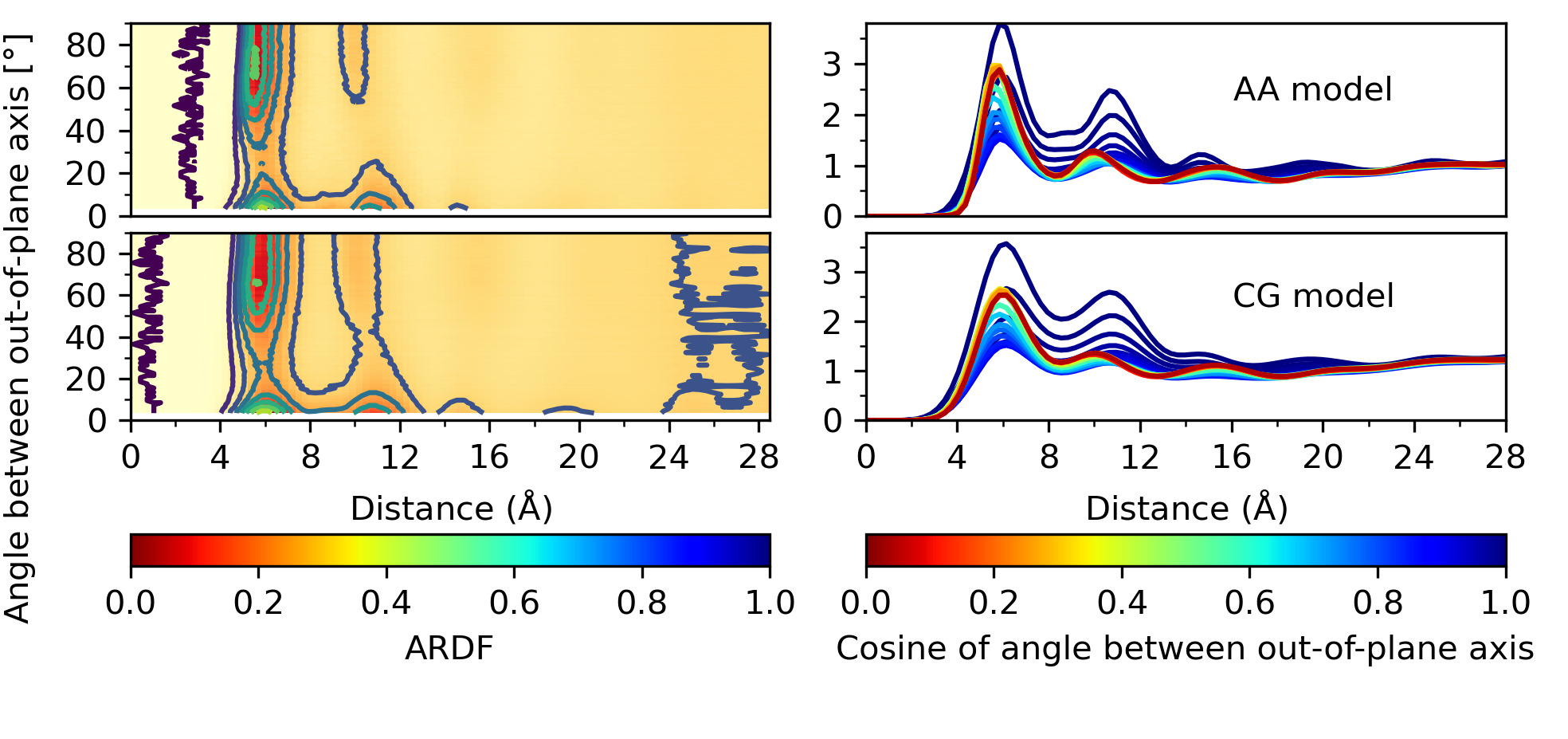}
	\caption{Angular-radial distribution function (ARDF) of the all-atom (AA) (top) and coarse-grained (CG) (bottom) sexithiophene models at 590~K and 1~atm (smectic phase) depicted as a heat map (left) and 1D slices at constant angle (right). Face-to-face, edge-to-edge, and parallel displaced configurations occur when the angle is 0\textdegree, while edge-to-face configurations occur at 90\textdegree.} 
	\label{fig:sexithiopheneArdf590}
\end{figure*}

The coarse-grained model also accurately describes orientational correlations in condensed-phase sexithiophene, as illustrated by a comparison with the angular-radial distribution function of the all-atom model. At the parameterization temperature, the coarse-grained model is able to capture all major features when compared to the all-atom model (Fig.~\ref{fig:sexithiopheneArdf}). The neural-network model is also able to capture the relevant features in the structure of sexithiophene's smectic liquid-crystal phase at 590~K, as shown in Fig.~\ref{fig:sexithiopheneArdf590}. The discrepancies in the width and height of the peaks are likely due to the differences in molecular shape away from the parameterization temperature that was mentioned earlier. The ARDFs of the two models in the nematic phase at 640~K are compared in the Supplementary Material and show similarly good agreement. 

Despite sexithiophene not strictly meeting the conditions to be coarse-grained to a single anisotropic particle due to its significant flexibility, the coarse-grained neural-network model is still able to reproduce its condensed-phase structural properties and phase behavior with remarkable accuracy. The limitation of the single-site model is only evident under conditions where the conformation of the molecule is highly temperature-dependent. One way to construct a neural network model that is independent of temperature would be to extract the training data from multiple temperatures and define the molecular dimensions as the average over the crystalline and isotropic phases. While the results for sexithiophene are substantially better than expected given its flexibility, improvements can be made to the model by considering a coarse-grained mapping consisting of more than one site.\cite{DAdamo2015}

The coarse-grained simulation of sexithiophene demonstrated a speed-up of $132 \times$ compared with the all-atom simulation using the same hardware employed for the benzene simulations. This speedup is primarily due to the large reduction in the number of degrees of freedom in coarse-graining this molecule.

\section*{Conclusions}

We have applied machine learning and a recently derived systematic coarse-graining method for anisotropic particles to develop a single-site anisotropic coarse-grained potential of a molecular system. The iterative training of the neural-network potential is able to reproduce the forces, torques, and pressure of the fine-grained all-atom system. The final loss of the iterative training model was identical to the loss obtained from k-fold cross-validation. The CG model performs well for a rigid molecule like benzene but, remarkably, it also describes the phase behavior and molecular-scale structural correlations of a flexible molecule like sexithiophene with comparable accuracy, even though the aspect ratio of the molecule changes significantly over the simulated temperature range. We have demonstrated the versatility of the coarse-graining method by parameterizing models of benzene and sexithiophene at a single temperature and then studying their accuracy in capturing the structural properties of the corresponding all-atom model at different temperatures.  The sexithiophene model was also used to show the ability of the model to reproduce the phase behavior of the all-atom model, with the lowest fidelity coming from the crystalline phase, where the aspect ratio of the molecule had the largest deviation from the parameterization data set. A natural extension to this work would be to generalize the method to a multi-site anisotropic coarse-grained model for flexible molecules and polymers. 

\section*{Author's Contributions}

MOW: Investigation, methodology, formal analysis, visualization, writing -- original draft. DMH: Conceptualization, methodology, supervision, writing -- review and editing.

\begin{acknowledgments}
This work was supported by the Australian Research Council under the Discovery Projects
funding scheme (DP190102100). This research was undertaken with the assistance of resources and services from the National Computational Infrastructure (NCI), which is supported by the Australian Government, and from the University of Adelaide’s Phoenix High-Performance Computing service. 
\end{acknowledgments}

\section*{Author Declarations}

\subsection*{Conflict of Interest}

The authors have no conflicts to disclose.

\section*{Data Availability}

The Supplementary Material provides more extensive structural comparisons between the all-atom and coarse-grained simulations, more details on the methods used for the calculations, and a list of the software needed to implement and train the neural-network potential, along with links to their GitHub repositories. Simulation files and neural network scripts can be found at https://doi.org/10.25909/21218057.v1


\section*{References}
\bibliography{afm-cg-nn}

\begin{thebibliography}{64}%
\makeatletter
\providecommand \@ifxundefined [1]{%
 \@ifx{#1\undefined}
}%
\providecommand \@ifnum [1]{%
 \ifnum #1\expandafter \@firstoftwo
 \else \expandafter \@secondoftwo
 \fi
}%
\providecommand \@ifx [1]{%
 \ifx #1\expandafter \@firstoftwo
 \else \expandafter \@secondoftwo
 \fi
}%
\providecommand \natexlab [1]{#1}%
\providecommand \enquote  [1]{``#1''}%
\providecommand \bibnamefont  [1]{#1}%
\providecommand \bibfnamefont [1]{#1}%
\providecommand \citenamefont [1]{#1}%
\providecommand \href@noop [0]{\@secondoftwo}%
\providecommand \href [0]{\begingroup \@sanitize@url \@href}%
\providecommand \@href[1]{\@@startlink{#1}\@@href}%
\providecommand \@@href[1]{\endgroup#1\@@endlink}%
\providecommand \@sanitize@url [0]{\catcode `\\12\catcode `\$12\catcode
  `\&12\catcode `\#12\catcode `\^12\catcode `\_12\catcode `\%12\relax}%
\providecommand \@@startlink[1]{}%
\providecommand \@@endlink[0]{}%
\providecommand \url  [0]{\begingroup\@sanitize@url \@url }%
\providecommand \@url [1]{\endgroup\@href {#1}{\urlprefix }}%
\providecommand \urlprefix  [0]{URL }%
\providecommand \Eprint [0]{\href }%
\providecommand \doibase [0]{http://dx.doi.org/}%
\providecommand \selectlanguage [0]{\@gobble}%
\providecommand \bibinfo  [0]{\@secondoftwo}%
\providecommand \bibfield  [0]{\@secondoftwo}%
\providecommand \translation [1]{[#1]}%
\providecommand \BibitemOpen [0]{}%
\providecommand \bibitemStop [0]{}%
\providecommand \bibitemNoStop [0]{.\EOS\space}%
\providecommand \EOS [0]{\spacefactor3000\relax}%
\providecommand \BibitemShut  [1]{\csname bibitem#1\endcsname}%
\let\auto@bib@innerbib\@empty
\bibitem [{\citenamefont {Butler}\ \emph {et~al.}(2018)\citenamefont {Butler},
  \citenamefont {Davies}, \citenamefont {Cartwright}, \citenamefont {Isayev},\
  and\ \citenamefont {Walsh}}]{butler2018machine}%
  \BibitemOpen
  \bibfield  {author} {\bibinfo {author} {\bibfnamefont {K.~T.}\ \bibnamefont
  {Butler}}, \bibinfo {author} {\bibfnamefont {D.~W.}\ \bibnamefont {Davies}},
  \bibinfo {author} {\bibfnamefont {H.}~\bibnamefont {Cartwright}}, \bibinfo
  {author} {\bibfnamefont {O.}~\bibnamefont {Isayev}}, \ and\ \bibinfo {author}
  {\bibfnamefont {A.}~\bibnamefont {Walsh}},\ }\bibfield  {title} {\enquote
  {\bibinfo {title} {Machine learning for molecular and materials science},}\
  }\href@noop {} {\bibfield  {journal} {\bibinfo  {journal} {Nature}\ }\textbf
  {\bibinfo {volume} {559}},\ \bibinfo {pages} {547--555} (\bibinfo {year}
  {2018})}\BibitemShut {NoStop}%
\bibitem [{\citenamefont {Moosavi}, \citenamefont {Jablonka},\ and\
  \citenamefont {Smit}(2020)}]{moosavi2020role}%
  \BibitemOpen
  \bibfield  {author} {\bibinfo {author} {\bibfnamefont {S.~M.}\ \bibnamefont
  {Moosavi}}, \bibinfo {author} {\bibfnamefont {K.~M.}\ \bibnamefont
  {Jablonka}}, \ and\ \bibinfo {author} {\bibfnamefont {B.}~\bibnamefont
  {Smit}},\ }\bibfield  {title} {\enquote {\bibinfo {title} {The role of
  machine learning in the understanding and design of materials},}\ }\href@noop
  {} {\bibfield  {journal} {\bibinfo  {journal} {J. Am. Chem. Soc.}\ }\textbf
  {\bibinfo {volume} {142}},\ \bibinfo {pages} {20273--20287} (\bibinfo {year}
  {2020})}\BibitemShut {NoStop}%
\bibitem [{\citenamefont {Tunyasuvunakool}\ \emph {et~al.}(2021)\citenamefont
  {Tunyasuvunakool}, \citenamefont {Adler}, \citenamefont {Wu}, \citenamefont
  {Green}, \citenamefont {Zielinski}, \citenamefont {{\v Z}{\'i}dek},
  \citenamefont {Bridgland}, \citenamefont {Cowie}, \citenamefont {Meyer},
  \citenamefont {Laydon}, \citenamefont {Velankar}, \citenamefont {Kleywegt},
  \citenamefont {Bateman}, \citenamefont {Evans}, \citenamefont {Pritzel},
  \citenamefont {Figurnov}, \citenamefont {Ronneberger}, \citenamefont {Bates},
  \citenamefont {Kohl}, \citenamefont {Potapenko}, \citenamefont {Ballard},
  \citenamefont {{Romera-Paredes}}, \citenamefont {Nikolov}, \citenamefont
  {Jain}, \citenamefont {Clancy}, \citenamefont {Reiman}, \citenamefont
  {Petersen}, \citenamefont {Senior}, \citenamefont {Kavukcuoglu},
  \citenamefont {Birney}, \citenamefont {Kohli}, \citenamefont {Jumper},\ and\
  \citenamefont {Hassabis}}]{tunyasuvunakool2021}%
  \BibitemOpen
  \bibfield  {author} {\bibinfo {author} {\bibfnamefont {K.}~\bibnamefont
  {Tunyasuvunakool}}, \bibinfo {author} {\bibfnamefont {J.}~\bibnamefont
  {Adler}}, \bibinfo {author} {\bibfnamefont {Z.}~\bibnamefont {Wu}}, \bibinfo
  {author} {\bibfnamefont {T.}~\bibnamefont {Green}}, \bibinfo {author}
  {\bibfnamefont {M.}~\bibnamefont {Zielinski}}, \bibinfo {author}
  {\bibfnamefont {A.}~\bibnamefont {{\v Z}{\'i}dek}}, \bibinfo {author}
  {\bibfnamefont {A.}~\bibnamefont {Bridgland}}, \bibinfo {author}
  {\bibfnamefont {A.}~\bibnamefont {Cowie}}, \bibinfo {author} {\bibfnamefont
  {C.}~\bibnamefont {Meyer}}, \bibinfo {author} {\bibfnamefont
  {A.}~\bibnamefont {Laydon}}, \bibinfo {author} {\bibfnamefont
  {S.}~\bibnamefont {Velankar}}, \bibinfo {author} {\bibfnamefont {G.~J.}\
  \bibnamefont {Kleywegt}}, \bibinfo {author} {\bibfnamefont {A.}~\bibnamefont
  {Bateman}}, \bibinfo {author} {\bibfnamefont {R.}~\bibnamefont {Evans}},
  \bibinfo {author} {\bibfnamefont {A.}~\bibnamefont {Pritzel}}, \bibinfo
  {author} {\bibfnamefont {M.}~\bibnamefont {Figurnov}}, \bibinfo {author}
  {\bibfnamefont {O.}~\bibnamefont {Ronneberger}}, \bibinfo {author}
  {\bibfnamefont {R.}~\bibnamefont {Bates}}, \bibinfo {author} {\bibfnamefont
  {S.~A.~A.}\ \bibnamefont {Kohl}}, \bibinfo {author} {\bibfnamefont
  {A.}~\bibnamefont {Potapenko}}, \bibinfo {author} {\bibfnamefont {A.~J.}\
  \bibnamefont {Ballard}}, \bibinfo {author} {\bibfnamefont {B.}~\bibnamefont
  {{Romera-Paredes}}}, \bibinfo {author} {\bibfnamefont {S.}~\bibnamefont
  {Nikolov}}, \bibinfo {author} {\bibfnamefont {R.}~\bibnamefont {Jain}},
  \bibinfo {author} {\bibfnamefont {E.}~\bibnamefont {Clancy}}, \bibinfo
  {author} {\bibfnamefont {D.}~\bibnamefont {Reiman}}, \bibinfo {author}
  {\bibfnamefont {S.}~\bibnamefont {Petersen}}, \bibinfo {author}
  {\bibfnamefont {A.~W.}\ \bibnamefont {Senior}}, \bibinfo {author}
  {\bibfnamefont {K.}~\bibnamefont {Kavukcuoglu}}, \bibinfo {author}
  {\bibfnamefont {E.}~\bibnamefont {Birney}}, \bibinfo {author} {\bibfnamefont
  {P.}~\bibnamefont {Kohli}}, \bibinfo {author} {\bibfnamefont
  {J.}~\bibnamefont {Jumper}}, \ and\ \bibinfo {author} {\bibfnamefont
  {D.}~\bibnamefont {Hassabis}},\ }\bibfield  {title} {\enquote {\bibinfo
  {title} {Highly accurate protein structure prediction for the human
  proteome},}\ }\href {\doibase 10.1038/s41586-021-03828-1} {\bibfield
  {journal} {\bibinfo  {journal} {Nature}\ }\textbf {\bibinfo {volume} {596}},\
  \bibinfo {pages} {590--596} (\bibinfo {year} {2021})}\BibitemShut {NoStop}%
\bibitem [{\citenamefont {Rowe}\ \emph {et~al.}(2020)\citenamefont {Rowe},
  \citenamefont {Deringer}, \citenamefont {Gasparotto}, \citenamefont
  {Cs{\'a}nyi},\ and\ \citenamefont {Michaelides}}]{rowe2020accurate}%
  \BibitemOpen
  \bibfield  {author} {\bibinfo {author} {\bibfnamefont {P.}~\bibnamefont
  {Rowe}}, \bibinfo {author} {\bibfnamefont {V.~L.}\ \bibnamefont {Deringer}},
  \bibinfo {author} {\bibfnamefont {P.}~\bibnamefont {Gasparotto}}, \bibinfo
  {author} {\bibfnamefont {G.}~\bibnamefont {Cs{\'a}nyi}}, \ and\ \bibinfo
  {author} {\bibfnamefont {A.}~\bibnamefont {Michaelides}},\ }\bibfield
  {title} {\enquote {\bibinfo {title} {An accurate and transferable machine
  learning potential for carbon},}\ }\href@noop {} {\bibfield  {journal}
  {\bibinfo  {journal} {J. Chem. Phys.}\ }\textbf {\bibinfo {volume} {153}},\
  \bibinfo {pages} {034702} (\bibinfo {year} {2020})}\BibitemShut {NoStop}%
\bibitem [{\citenamefont {Stocker}\ \emph {et~al.}(2022)\citenamefont
  {Stocker}, \citenamefont {Gasteiger}, \citenamefont {Becker}, \citenamefont
  {G{\"u}nnemann},\ and\ \citenamefont {Margraf}}]{stocker2022robust}%
  \BibitemOpen
  \bibfield  {author} {\bibinfo {author} {\bibfnamefont {S.}~\bibnamefont
  {Stocker}}, \bibinfo {author} {\bibfnamefont {J.}~\bibnamefont {Gasteiger}},
  \bibinfo {author} {\bibfnamefont {F.}~\bibnamefont {Becker}}, \bibinfo
  {author} {\bibfnamefont {S.}~\bibnamefont {G{\"u}nnemann}}, \ and\ \bibinfo
  {author} {\bibfnamefont {J.~T.}\ \bibnamefont {Margraf}},\ }\bibfield
  {title} {\enquote {\bibinfo {title} {How robust are modern graph neural
  network potentials in long and hot molecular dynamics simulations?}}\
  }\href@noop {} {\bibfield  {journal} {\bibinfo  {journal} {Mach. Learn.: Sci.
  Tech.}\ }\textbf {\bibinfo {volume} {3}},\ \bibinfo {pages} {045010}
  (\bibinfo {year} {2022})}\BibitemShut {NoStop}%
\bibitem [{\citenamefont {Friederich}\ \emph {et~al.}(2021)\citenamefont
  {Friederich}, \citenamefont {H{\"a}se}, \citenamefont {Proppe},\ and\
  \citenamefont {Aspuru-Guzik}}]{friederich2021machine}%
  \BibitemOpen
  \bibfield  {author} {\bibinfo {author} {\bibfnamefont {P.}~\bibnamefont
  {Friederich}}, \bibinfo {author} {\bibfnamefont {F.}~\bibnamefont
  {H{\"a}se}}, \bibinfo {author} {\bibfnamefont {J.}~\bibnamefont {Proppe}}, \
  and\ \bibinfo {author} {\bibfnamefont {A.}~\bibnamefont {Aspuru-Guzik}},\
  }\bibfield  {title} {\enquote {\bibinfo {title} {Machine-learned potentials
  for next-generation matter simulations},}\ }\href@noop {} {\bibfield
  {journal} {\bibinfo  {journal} {Nat. Mater.}\ }\textbf {\bibinfo {volume}
  {20}},\ \bibinfo {pages} {750--761} (\bibinfo {year} {2021})}\BibitemShut
  {NoStop}%
\bibitem [{\citenamefont {No{\'e}}\ \emph {et~al.}(2020)\citenamefont
  {No{\'e}}, \citenamefont {Tkatchenko}, \citenamefont {M{\"u}ller},\ and\
  \citenamefont {Clementi}}]{noe2020machine}%
  \BibitemOpen
  \bibfield  {author} {\bibinfo {author} {\bibfnamefont {F.}~\bibnamefont
  {No{\'e}}}, \bibinfo {author} {\bibfnamefont {A.}~\bibnamefont {Tkatchenko}},
  \bibinfo {author} {\bibfnamefont {K.-R.}\ \bibnamefont {M{\"u}ller}}, \ and\
  \bibinfo {author} {\bibfnamefont {C.}~\bibnamefont {Clementi}},\ }\bibfield
  {title} {\enquote {\bibinfo {title} {Machine learning for molecular
  simulation},}\ }\href@noop {} {\bibfield  {journal} {\bibinfo  {journal}
  {Annu. Rev. Phys. Chem.}\ }\textbf {\bibinfo {volume} {71}},\ \bibinfo
  {pages} {361--390} (\bibinfo {year} {2020})}\BibitemShut {NoStop}%
\bibitem [{\citenamefont {Guo}\ \emph {et~al.}(2022)\citenamefont {Guo},
  \citenamefont {Lu}, \citenamefont {Yan}, \citenamefont {Hu}, \citenamefont
  {Liu}, \citenamefont {Tan}, \citenamefont {Sun}, \citenamefont {Jiang},
  \citenamefont {Liu}, \citenamefont {Chen}, \citenamefont {Zhang},
  \citenamefont {Chen}, \citenamefont {Wang},\ and\ \citenamefont
  {Jia}}]{Guo2022}%
  \BibitemOpen
  \bibfield  {author} {\bibinfo {author} {\bibfnamefont {Z.}~\bibnamefont
  {Guo}}, \bibinfo {author} {\bibfnamefont {D.}~\bibnamefont {Lu}}, \bibinfo
  {author} {\bibfnamefont {Y.}~\bibnamefont {Yan}}, \bibinfo {author}
  {\bibfnamefont {S.}~\bibnamefont {Hu}}, \bibinfo {author} {\bibfnamefont
  {R.}~\bibnamefont {Liu}}, \bibinfo {author} {\bibfnamefont {G.}~\bibnamefont
  {Tan}}, \bibinfo {author} {\bibfnamefont {N.}~\bibnamefont {Sun}}, \bibinfo
  {author} {\bibfnamefont {W.}~\bibnamefont {Jiang}}, \bibinfo {author}
  {\bibfnamefont {L.}~\bibnamefont {Liu}}, \bibinfo {author} {\bibfnamefont
  {Y.}~\bibnamefont {Chen}}, \bibinfo {author} {\bibfnamefont {L.}~\bibnamefont
  {Zhang}}, \bibinfo {author} {\bibfnamefont {M.}~\bibnamefont {Chen}},
  \bibinfo {author} {\bibfnamefont {H.}~\bibnamefont {Wang}}, \ and\ \bibinfo
  {author} {\bibfnamefont {W.}~\bibnamefont {Jia}},\ }\bibfield  {title}
  {\enquote {\bibinfo {title} {Extending the limit of molecular dynamics with
  ab initio accuracy to 10 billion atoms},}\ }in\ \href {\doibase
  10.1145/3503221.3508425} {\emph {\bibinfo {booktitle} {Proc. 27th {{ACM
  SIGPLAN Symp}}. {{Princ}}. {{Pract}}. {{Parallel Program}}.}}},\ \bibinfo
  {series and number} {{{PPoPP}} '22}\ (\bibinfo  {publisher} {{Association for
  Computing Machinery}},\ \bibinfo {address} {{New York, NY, USA}},\ \bibinfo
  {year} {2022})\ pp.\ \bibinfo {pages} {205--218}\BibitemShut {NoStop}%
\bibitem [{\citenamefont {Marx}\ and\ \citenamefont {Hutter}(2009)}]{Marx2009}%
  \BibitemOpen
  \bibfield  {author} {\bibinfo {author} {\bibfnamefont {D.}~\bibnamefont
  {Marx}}\ and\ \bibinfo {author} {\bibfnamefont {J.}~\bibnamefont {Hutter}},\
  }\href@noop {} {\emph {\bibinfo {title} {Ab initio molecular dynamics: basic
  theory and advanced methods}}}\ (\bibinfo  {publisher} {{Cambridge University
  Press}},\ \bibinfo {year} {2009})\BibitemShut {NoStop}%
\bibitem [{\citenamefont {Wang}\ \emph {et~al.}(2018)\citenamefont {Wang},
  \citenamefont {Zhang}, \citenamefont {Han},\ and\ \citenamefont
  {E}}]{Wang2018}%
  \BibitemOpen
  \bibfield  {author} {\bibinfo {author} {\bibfnamefont {H.}~\bibnamefont
  {Wang}}, \bibinfo {author} {\bibfnamefont {L.}~\bibnamefont {Zhang}},
  \bibinfo {author} {\bibfnamefont {J.}~\bibnamefont {Han}}, \ and\ \bibinfo
  {author} {\bibfnamefont {W.}~\bibnamefont {E}},\ }\bibfield  {title}
  {\enquote {\bibinfo {title} {{{DeePMD-kit}}: {{A}} deep learning package for
  many-body potential energy representation and molecular dynamics},}\ }\href
  {\doibase 10.1016/j.cpc.2018.03.016} {\bibfield  {journal} {\bibinfo
  {journal} {Comput. Phys. Commun.}\ }\textbf {\bibinfo {volume} {228}},\
  \bibinfo {pages} {178--184} (\bibinfo {year} {2018})}\BibitemShut {NoStop}%
\bibitem [{\citenamefont {Unke}\ \emph {et~al.}(2021)\citenamefont {Unke},
  \citenamefont {Chmiela}, \citenamefont {Sauceda}, \citenamefont {Gastegger},
  \citenamefont {Poltavsky}, \citenamefont {Sch{\"u}tt}, \citenamefont
  {Tkatchenko},\ and\ \citenamefont {M{\"u}ller}}]{unke2021machine}%
  \BibitemOpen
  \bibfield  {author} {\bibinfo {author} {\bibfnamefont {O.~T.}\ \bibnamefont
  {Unke}}, \bibinfo {author} {\bibfnamefont {S.}~\bibnamefont {Chmiela}},
  \bibinfo {author} {\bibfnamefont {H.~E.}\ \bibnamefont {Sauceda}}, \bibinfo
  {author} {\bibfnamefont {M.}~\bibnamefont {Gastegger}}, \bibinfo {author}
  {\bibfnamefont {I.}~\bibnamefont {Poltavsky}}, \bibinfo {author}
  {\bibfnamefont {K.~T.}\ \bibnamefont {Sch{\"u}tt}}, \bibinfo {author}
  {\bibfnamefont {A.}~\bibnamefont {Tkatchenko}}, \ and\ \bibinfo {author}
  {\bibfnamefont {K.-R.}\ \bibnamefont {M{\"u}ller}},\ }\bibfield  {title}
  {\enquote {\bibinfo {title} {Machine learning force fields},}\ }\href@noop {}
  {\bibfield  {journal} {\bibinfo  {journal} {Chem. Rev.}\ }\textbf {\bibinfo
  {volume} {121}},\ \bibinfo {pages} {10142--10186} (\bibinfo {year}
  {2021})}\BibitemShut {NoStop}%
\bibitem [{\citenamefont {Jin}\ \emph {et~al.}(2022)\citenamefont {Jin},
  \citenamefont {Pak}, \citenamefont {Durumeric}, \citenamefont {Loose},\ and\
  \citenamefont {Voth}}]{jin2022bottom}%
  \BibitemOpen
  \bibfield  {author} {\bibinfo {author} {\bibfnamefont {J.}~\bibnamefont
  {Jin}}, \bibinfo {author} {\bibfnamefont {A.~J.}\ \bibnamefont {Pak}},
  \bibinfo {author} {\bibfnamefont {A.~E.}\ \bibnamefont {Durumeric}}, \bibinfo
  {author} {\bibfnamefont {T.~D.}\ \bibnamefont {Loose}}, \ and\ \bibinfo
  {author} {\bibfnamefont {G.~A.}\ \bibnamefont {Voth}},\ }\bibfield  {title}
  {\enquote {\bibinfo {title} {Bottom-up coarse-graining: Principles and
  perspectives},}\ }\href@noop {} {\bibfield  {journal} {\bibinfo  {journal}
  {J. Chem. Theory Comput.}\ }\textbf {\bibinfo {volume} {18}},\ \bibinfo
  {pages} {5759--5791} (\bibinfo {year} {2022})}\BibitemShut {NoStop}%
\bibitem [{\citenamefont {Marrink}\ and\ \citenamefont
  {Tieleman}(2013)}]{marrink2013perspective}%
  \BibitemOpen
  \bibfield  {author} {\bibinfo {author} {\bibfnamefont {S.~J.}\ \bibnamefont
  {Marrink}}\ and\ \bibinfo {author} {\bibfnamefont {D.~P.}\ \bibnamefont
  {Tieleman}},\ }\bibfield  {title} {\enquote {\bibinfo {title} {Perspective on
  the {M}artini model},}\ }\href@noop {} {\bibfield  {journal} {\bibinfo
  {journal} {Chem. Soc. Rev.}\ }\textbf {\bibinfo {volume} {42}},\ \bibinfo
  {pages} {6801--6822} (\bibinfo {year} {2013})}\BibitemShut {NoStop}%
\bibitem [{\citenamefont {Zhang}\ \emph {et~al.}(2018)\citenamefont {Zhang},
  \citenamefont {Han}, \citenamefont {Wang}, \citenamefont {Car},\ and\
  \citenamefont {Weinan}}]{Zhang2018}%
  \BibitemOpen
  \bibfield  {author} {\bibinfo {author} {\bibfnamefont {L.}~\bibnamefont
  {Zhang}}, \bibinfo {author} {\bibfnamefont {J.}~\bibnamefont {Han}}, \bibinfo
  {author} {\bibfnamefont {H.}~\bibnamefont {Wang}}, \bibinfo {author}
  {\bibfnamefont {R.}~\bibnamefont {Car}}, \ and\ \bibinfo {author}
  {\bibfnamefont {W.~E.}\ \bibnamefont {Weinan}},\ }\bibfield  {title}
  {\enquote {\bibinfo {title} {{{DeePCG}}: {{Constructing}} coarse-grained
  models via deep neural networks},}\ }\href {\doibase 10.1063/1.5027645}
  {\bibfield  {journal} {\bibinfo  {journal} {J. Chem. Phys.}\ }\textbf
  {\bibinfo {volume} {149}},\ \bibinfo {pages} {034101} (\bibinfo {year}
  {2018})}\BibitemShut {NoStop}%
\bibitem [{\citenamefont {Wang}\ \emph {et~al.}(2019)\citenamefont {Wang},
  \citenamefont {Olsson}, \citenamefont {Wehmeyer}, \citenamefont {P{\'e}rez},
  \citenamefont {Charron}, \citenamefont {Fabritiis}, \citenamefont {No{\'e}},\
  and\ \citenamefont {Clementi}}]{Wang2019}%
  \BibitemOpen
  \bibfield  {author} {\bibinfo {author} {\bibfnamefont {J.}~\bibnamefont
  {Wang}}, \bibinfo {author} {\bibfnamefont {S.}~\bibnamefont {Olsson}},
  \bibinfo {author} {\bibfnamefont {C.}~\bibnamefont {Wehmeyer}}, \bibinfo
  {author} {\bibfnamefont {A.}~\bibnamefont {P{\'e}rez}}, \bibinfo {author}
  {\bibfnamefont {N.~E.}\ \bibnamefont {Charron}}, \bibinfo {author}
  {\bibfnamefont {G.~D.}\ \bibnamefont {Fabritiis}}, \bibinfo {author}
  {\bibfnamefont {F.}~\bibnamefont {No{\'e}}}, \ and\ \bibinfo {author}
  {\bibfnamefont {C.}~\bibnamefont {Clementi}},\ }\bibfield  {title} {\enquote
  {\bibinfo {title} {Machine learning of coarse-grained molecular dynamics
  force fields},}\ }\href {\doibase 10.1021/acscentsci.8b00913} {\bibfield
  {journal} {\bibinfo  {journal} {ACS Cent. Sci.}\ }\textbf {\bibinfo {volume}
  {5}},\ \bibinfo {pages} {755--767} (\bibinfo {year} {2019})}\BibitemShut
  {NoStop}%
\bibitem [{\citenamefont {Gay}\ and\ \citenamefont {Berne}(1981)}]{Gay1981}%
  \BibitemOpen
  \bibfield  {author} {\bibinfo {author} {\bibfnamefont {J.~G.}\ \bibnamefont
  {Gay}}\ and\ \bibinfo {author} {\bibfnamefont {B.~J.}\ \bibnamefont
  {Berne}},\ }\bibfield  {title} {\enquote {\bibinfo {title} {Modification of
  the overlap potential to mimic a linear site\textendash site potential},}\
  }\href {\doibase 10.1063/1.441483} {\bibfield  {journal} {\bibinfo  {journal}
  {J. Chem. Phys.}\ }\textbf {\bibinfo {volume} {74}},\ \bibinfo {pages}
  {3316--3319} (\bibinfo {year} {1981})}\BibitemShut {NoStop}%
\bibitem [{\citenamefont {Berardi}, \citenamefont {Fava},\ and\ \citenamefont
  {Zannoni}(1995)}]{Berardi1995}%
  \BibitemOpen
  \bibfield  {author} {\bibinfo {author} {\bibfnamefont {R.}~\bibnamefont
  {Berardi}}, \bibinfo {author} {\bibfnamefont {C.}~\bibnamefont {Fava}}, \
  and\ \bibinfo {author} {\bibfnamefont {C.}~\bibnamefont {Zannoni}},\
  }\bibfield  {title} {\enquote {\bibinfo {title} {A generalized
  {{Gay–Berne}} intermolecular potential for biaxial particles},}\ }\href
  {\doibase 10.1016/0009-2614(95)00212-M} {\bibfield  {journal} {\bibinfo
  {journal} {Chem. Phys. Lett.}\ }\textbf {\bibinfo {volume} {236}},\ \bibinfo
  {pages} {462--468} (\bibinfo {year} {1995})}\BibitemShut {NoStop}%
\bibitem [{\citenamefont {Boehm}, \citenamefont {Nguyen},\ and\ \citenamefont
  {Huang}(2019)}]{boehm2019interplay}%
  \BibitemOpen
  \bibfield  {author} {\bibinfo {author} {\bibfnamefont {B.~J.}\ \bibnamefont
  {Boehm}}, \bibinfo {author} {\bibfnamefont {H.~T.}\ \bibnamefont {Nguyen}}, \
  and\ \bibinfo {author} {\bibfnamefont {D.~M.}\ \bibnamefont {Huang}},\
  }\bibfield  {title} {\enquote {\bibinfo {title} {The interplay of interfaces,
  supramolecular assembly, and electronics in organic semiconductors},}\
  }\href@noop {} {\bibfield  {journal} {\bibinfo  {journal} {J. Phys.: Condens.
  Matter}\ }\textbf {\bibinfo {volume} {31}},\ \bibinfo {pages} {423001}
  (\bibinfo {year} {2019})}\BibitemShut {NoStop}%
\bibitem [{\citenamefont {Behler}(2011)}]{Behler2011}%
  \BibitemOpen
  \bibfield  {author} {\bibinfo {author} {\bibfnamefont {J.}~\bibnamefont
  {Behler}},\ }\bibfield  {title} {\enquote {\bibinfo {title} {Atom-centered
  symmetry functions for constructing high-dimensional neural network
  potentials},}\ }\href {\doibase 10.1063/1.3553717} {\bibfield  {journal}
  {\bibinfo  {journal} {J. Chem. Phys.}\ }\textbf {\bibinfo {volume} {134}},\
  \bibinfo {pages} {074106} (\bibinfo {year} {2011})}\BibitemShut {NoStop}%
\bibitem [{\citenamefont {{Campos-Villalobos}}\ \emph
  {et~al.}(2022)\citenamefont {{Campos-Villalobos}}, \citenamefont {Giunta},
  \citenamefont {{Mar{\'i}n-Aguilar}},\ and\ \citenamefont
  {Dijkstra}}]{CamposVillalobos2022}%
  \BibitemOpen
  \bibfield  {author} {\bibinfo {author} {\bibfnamefont {G.}~\bibnamefont
  {{Campos-Villalobos}}}, \bibinfo {author} {\bibfnamefont {G.}~\bibnamefont
  {Giunta}}, \bibinfo {author} {\bibfnamefont {S.}~\bibnamefont
  {{Mar{\'i}n-Aguilar}}}, \ and\ \bibinfo {author} {\bibfnamefont
  {M.}~\bibnamefont {Dijkstra}},\ }\bibfield  {title} {\enquote {\bibinfo
  {title} {Machine-learning effective many-body potentials for anisotropic
  particles using orientation-dependent symmetry functions},}\ }\href@noop {}
  {\bibfield  {journal} {\bibinfo  {journal} {J. Chem. Phys.}\ }\textbf
  {\bibinfo {volume} {157}},\ \bibinfo {pages} {024902} (\bibinfo {year}
  {2022})}\BibitemShut {NoStop}%
\bibitem [{\citenamefont {Nguyen}\ and\ \citenamefont
  {Huang}(2022)}]{Nguyen2022}%
  \BibitemOpen
  \bibfield  {author} {\bibinfo {author} {\bibfnamefont {H.~T.~L.}\
  \bibnamefont {Nguyen}}\ and\ \bibinfo {author} {\bibfnamefont {D.~M.}\
  \bibnamefont {Huang}},\ }\bibfield  {title} {\enquote {\bibinfo {title}
  {Systematic bottom-up molecular coarse-graining via force and torque matching
  using anisotropic particles},}\ }\href {\doibase 10.1063/5.0085006}
  {\bibfield  {journal} {\bibinfo  {journal} {J. Chem. Phys.}\ }\textbf
  {\bibinfo {volume} {156}},\ \bibinfo {pages} {184118} (\bibinfo {year}
  {2022})}\BibitemShut {NoStop}%
\bibitem [{\citenamefont {Noid}\ \emph {et~al.}(2008)\citenamefont {Noid},
  \citenamefont {Chu}, \citenamefont {Ayton}, \citenamefont {Krishna},
  \citenamefont {Izvekov}, \citenamefont {Voth}, \citenamefont {Das},\ and\
  \citenamefont {Andersen}}]{Noid2008}%
  \BibitemOpen
  \bibfield  {author} {\bibinfo {author} {\bibfnamefont {W.~G.}\ \bibnamefont
  {Noid}}, \bibinfo {author} {\bibfnamefont {J.-W.}\ \bibnamefont {Chu}},
  \bibinfo {author} {\bibfnamefont {G.~S.}\ \bibnamefont {Ayton}}, \bibinfo
  {author} {\bibfnamefont {V.}~\bibnamefont {Krishna}}, \bibinfo {author}
  {\bibfnamefont {S.}~\bibnamefont {Izvekov}}, \bibinfo {author} {\bibfnamefont
  {G.~A.}\ \bibnamefont {Voth}}, \bibinfo {author} {\bibfnamefont
  {A.}~\bibnamefont {Das}}, \ and\ \bibinfo {author} {\bibfnamefont {H.~C.}\
  \bibnamefont {Andersen}},\ }\bibfield  {title} {\enquote {\bibinfo {title}
  {The multiscale coarse-graining method. {{I}}. {{A}} rigorous bridge between
  atomistic and coarse-grained models},}\ }\href {\doibase 10.1063/1.2938860}
  {\bibfield  {journal} {\bibinfo  {journal} {J. Chem. Phys.}\ }\textbf
  {\bibinfo {volume} {128}},\ \bibinfo {pages} {244114} (\bibinfo {year}
  {2008})}\BibitemShut {NoStop}%
\bibitem [{\citenamefont {Das}\ and\ \citenamefont
  {Andersen}(2010)}]{das2010a}%
  \BibitemOpen
  \bibfield  {author} {\bibinfo {author} {\bibfnamefont {A.}~\bibnamefont
  {Das}}\ and\ \bibinfo {author} {\bibfnamefont {H.~C.}\ \bibnamefont
  {Andersen}},\ }\bibfield  {title} {\enquote {\bibinfo {title} {The multiscale
  coarse-graining method. {{V}}. {{Isothermal-isobaric}} ensemble},}\
  }\href@noop {} {\bibfield  {journal} {\bibinfo  {journal} {J. Chem. Phys.}\
  }\textbf {\bibinfo {volume} {132}},\ \bibinfo {pages} {164106} (\bibinfo
  {year} {2010})}\BibitemShut {NoStop}%
\bibitem [{\citenamefont {Goldstein}(2002)}]{Goldstein2002}%
  \BibitemOpen
  \bibfield  {author} {\bibinfo {author} {\bibfnamefont {H.}~\bibnamefont
  {Goldstein}},\ }\href@noop {} {\emph {\bibinfo {title} {Classical
  Mechanics}}}\ (\bibinfo  {publisher} {Addison-Wesley San Francisco},\
  \bibinfo {year} {2002})\BibitemShut {NoStop}%
\bibitem [{\citenamefont {Bart{\'o}k}, \citenamefont {Kondor},\ and\
  \citenamefont {Cs{\'a}nyi}(2013)}]{Bartok2013}%
  \BibitemOpen
  \bibfield  {author} {\bibinfo {author} {\bibfnamefont {A.~P.}\ \bibnamefont
  {Bart{\'o}k}}, \bibinfo {author} {\bibfnamefont {R.}~\bibnamefont {Kondor}},
  \ and\ \bibinfo {author} {\bibfnamefont {G.}~\bibnamefont {Cs{\'a}nyi}},\
  }\bibfield  {title} {\enquote {\bibinfo {title} {On representing chemical
  environments},}\ }\href {\doibase 10.1103/PhysRevB.87.184115} {\bibfield
  {journal} {\bibinfo  {journal} {Phys. Rev. B}\ }\textbf {\bibinfo {volume}
  {87}},\ \bibinfo {pages} {184115} (\bibinfo {year} {2013})}\BibitemShut
  {NoStop}%
\bibitem [{\citenamefont {Han}\ \emph {et~al.}(2018)\citenamefont {Han},
  \citenamefont {Zhang}, \citenamefont {Car},\ and\ \citenamefont
  {E}}]{Han2018}%
  \BibitemOpen
  \bibfield  {author} {\bibinfo {author} {\bibfnamefont {J.}~\bibnamefont
  {Han}}, \bibinfo {author} {\bibfnamefont {L.}~\bibnamefont {Zhang}}, \bibinfo
  {author} {\bibfnamefont {R.}~\bibnamefont {Car}}, \ and\ \bibinfo {author}
  {\bibfnamefont {W.}~\bibnamefont {E}},\ }\bibfield  {title} {\enquote
  {\bibinfo {title} {Deep {P}otential: {{A}} general representation of a
  many-body potential energy surface},}\ }\href {\doibase
  10.4208/cicp.OA-2017-0213} {\bibfield  {journal} {\bibinfo  {journal}
  {Commun. Comput. Phys.}\ }\textbf {\bibinfo {volume} {23}},\ \bibinfo {pages}
  {629--639} (\bibinfo {year} {2018})}\BibitemShut {NoStop}%
\bibitem [{\citenamefont {Zhou}\ \emph {et~al.}(2019)\citenamefont {Zhou},
  \citenamefont {Barnes}, \citenamefont {Lu}, \citenamefont {Yang},\ and\
  \citenamefont {Li}}]{Zhou2019}%
  \BibitemOpen
  \bibfield  {author} {\bibinfo {author} {\bibfnamefont {Y.}~\bibnamefont
  {Zhou}}, \bibinfo {author} {\bibfnamefont {C.}~\bibnamefont {Barnes}},
  \bibinfo {author} {\bibfnamefont {J.}~\bibnamefont {Lu}}, \bibinfo {author}
  {\bibfnamefont {J.}~\bibnamefont {Yang}}, \ and\ \bibinfo {author}
  {\bibfnamefont {H.}~\bibnamefont {Li}},\ }\bibfield  {title} {\enquote
  {\bibinfo {title} {On the continuity of rotation representations in neural
  networks},}\ }in\ \href {\doibase 10.1109/CVPR.2019.00589} {\emph {\bibinfo
  {booktitle} {2019 {{IEEECVF Conf}}. {{Comput}}. {{Vis}}. {{Pattern
  Recognit}}. {{CVPR}}}}}\ (\bibinfo  {publisher} {{IEEE}},\ \bibinfo {year}
  {2019})\ pp.\ \bibinfo {pages} {5738--5746}\BibitemShut {NoStop}%
\bibitem [{\citenamefont {Gastegger}\ \emph {et~al.}(2018)\citenamefont
  {Gastegger}, \citenamefont {Schwiedrzik}, \citenamefont {Bittermann},
  \citenamefont {Berzsenyi},\ and\ \citenamefont {Marquetand}}]{Gastegger2018}%
  \BibitemOpen
  \bibfield  {author} {\bibinfo {author} {\bibfnamefont {M.}~\bibnamefont
  {Gastegger}}, \bibinfo {author} {\bibfnamefont {L.}~\bibnamefont
  {Schwiedrzik}}, \bibinfo {author} {\bibfnamefont {M.}~\bibnamefont
  {Bittermann}}, \bibinfo {author} {\bibfnamefont {F.}~\bibnamefont
  {Berzsenyi}}, \ and\ \bibinfo {author} {\bibfnamefont {P.}~\bibnamefont
  {Marquetand}},\ }\bibfield  {title} {\enquote {\bibinfo {title}
  {{{wACSF}}\textemdash{{Weighted}} atom-centered symmetry functions as
  descriptors in machine learning potentials},}\ }\href {\doibase
  10.1063/1.5019667} {\bibfield  {journal} {\bibinfo  {journal} {J. Chem.
  Phys.}\ }\textbf {\bibinfo {volume} {148}},\ \bibinfo {pages} {241709}
  (\bibinfo {year} {2018})}\BibitemShut {NoStop}%
\bibitem [{\citenamefont {Ciccotti}, \citenamefont {Kapral},\ and\
  \citenamefont {{Vanden-Eijnden}}(2005)}]{Ciccotti2005}%
  \BibitemOpen
  \bibfield  {author} {\bibinfo {author} {\bibfnamefont {G.}~\bibnamefont
  {Ciccotti}}, \bibinfo {author} {\bibfnamefont {R.}~\bibnamefont {Kapral}}, \
  and\ \bibinfo {author} {\bibfnamefont {E.}~\bibnamefont {{Vanden-Eijnden}}},\
  }\bibfield  {title} {\enquote {\bibinfo {title} {Blue moon sampling,
  vectorial reaction coordinates, and unbiased constrained dynamics},}\ }\href
  {\doibase 10.1002/CPHC.200400669} {\bibfield  {journal} {\bibinfo  {journal}
  {ChemPhysChem}\ }\textbf {\bibinfo {volume} {6}},\ \bibinfo {pages}
  {1809--1814} (\bibinfo {year} {2005})}\BibitemShut {NoStop}%
\bibitem [{\citenamefont {Abrams}\ and\ \citenamefont
  {Tuckerman}(2008)}]{Abrams2008}%
  \BibitemOpen
  \bibfield  {author} {\bibinfo {author} {\bibfnamefont {J.~B.}\ \bibnamefont
  {Abrams}}\ and\ \bibinfo {author} {\bibfnamefont {M.~E.}\ \bibnamefont
  {Tuckerman}},\ }\bibfield  {title} {\enquote {\bibinfo {title} {Efficient and
  direct generation of multidimensional free energy surfaces via adiabatic
  dynamics without coordinate transformations},}\ }\href {\doibase
  10.1021/jp805039u} {\bibfield  {journal} {\bibinfo  {journal} {J. Phys. Chem.
  B}\ }\textbf {\bibinfo {volume} {112}},\ \bibinfo {pages} {15742--15757}
  (\bibinfo {year} {2008})}\BibitemShut {NoStop}%
\bibitem [{\citenamefont {Kingma}\ and\ \citenamefont {Ba}(2014)}]{Kingma2014}%
  \BibitemOpen
  \bibfield  {author} {\bibinfo {author} {\bibfnamefont {D.~P.}\ \bibnamefont
  {Kingma}}\ and\ \bibinfo {author} {\bibfnamefont {J.}~\bibnamefont {Ba}},\
  }\bibfield  {title} {\enquote {\bibinfo {title} {Adam: {{A}} method for
  stochastic optimization},}\ }\href@noop {} {\bibfield  {journal} {\bibinfo
  {journal} {arXiv}\ ,\ \bibinfo {pages} {1412.6980}} (\bibinfo {year}
  {2014})}\BibitemShut {NoStop}%
\bibitem [{\citenamefont {Abadi}\ \emph {et~al.}(2016)\citenamefont {Abadi},
  \citenamefont {Barham}, \citenamefont {Chen}, \citenamefont {Chen},
  \citenamefont {Davis}, \citenamefont {Dean}, \citenamefont {Devin},
  \citenamefont {Ghemawat}, \citenamefont {Irving}, \citenamefont {Isard} \emph
  {et~al.}}]{Abadi2016}%
  \BibitemOpen
  \bibfield  {author} {\bibinfo {author} {\bibfnamefont {M.}~\bibnamefont
  {Abadi}}, \bibinfo {author} {\bibfnamefont {P.}~\bibnamefont {Barham}},
  \bibinfo {author} {\bibfnamefont {J.}~\bibnamefont {Chen}}, \bibinfo {author}
  {\bibfnamefont {Z.}~\bibnamefont {Chen}}, \bibinfo {author} {\bibfnamefont
  {A.}~\bibnamefont {Davis}}, \bibinfo {author} {\bibfnamefont
  {J.}~\bibnamefont {Dean}}, \bibinfo {author} {\bibfnamefont {M.}~\bibnamefont
  {Devin}}, \bibinfo {author} {\bibfnamefont {S.}~\bibnamefont {Ghemawat}},
  \bibinfo {author} {\bibfnamefont {G.}~\bibnamefont {Irving}}, \bibinfo
  {author} {\bibfnamefont {M.}~\bibnamefont {Isard}},  \emph {et~al.},\
  }\bibfield  {title} {\enquote {\bibinfo {title} {{{TensorFlow}}: {{A}} system
  for large-scale machine learning},}\ }in\ \href@noop {} {\emph {\bibinfo
  {booktitle} {12th {{USENIX Symp}}. {{Oper}}. {{Syst}}. {{Des}}.
  {{Implement}}. {{OSDI}} 16}}}\ (\bibinfo {year} {2016})\ pp.\ \bibinfo
  {pages} {265--283}\BibitemShut {NoStop}%
\bibitem [{\citenamefont {Chollet}\ \emph {et~al.}(2015)\citenamefont {Chollet}
  \emph {et~al.}}]{Chollet2015}%
  \BibitemOpen
  \bibfield  {author} {\bibinfo {author} {\bibfnamefont {F.}~\bibnamefont
  {Chollet}} \emph {et~al.},\ }\href@noop {} {\enquote {\bibinfo {title}
  {Keras},}\ }\bibinfo {howpublished} {https://github.com/fchollet/keras}
  (\bibinfo {year} {2015})\BibitemShut {NoStop}%
\bibitem [{\citenamefont {Plimpton}(1995)}]{Plimpton1995}%
  \BibitemOpen
  \bibfield  {author} {\bibinfo {author} {\bibfnamefont {S.}~\bibnamefont
  {Plimpton}},\ }\bibfield  {title} {\enquote {\bibinfo {title} {Fast parallel
  algorithms for short-range molecular dynamics},}\ }\href@noop {} {\bibfield
  {journal} {\bibinfo  {journal} {J. Comput. Phys.}\ }\textbf {\bibinfo
  {volume} {117}},\ \bibinfo {pages} {1--19} (\bibinfo {year}
  {1995})}\BibitemShut {NoStop}%
\bibitem [{\citenamefont {Brown}\ \emph {et~al.}(2011)\citenamefont {Brown},
  \citenamefont {Wang}, \citenamefont {Plimpton},\ and\ \citenamefont
  {Tharrington}}]{Brown2011}%
  \BibitemOpen
  \bibfield  {author} {\bibinfo {author} {\bibfnamefont {W.~M.}\ \bibnamefont
  {Brown}}, \bibinfo {author} {\bibfnamefont {P.}~\bibnamefont {Wang}},
  \bibinfo {author} {\bibfnamefont {S.~J.}\ \bibnamefont {Plimpton}}, \ and\
  \bibinfo {author} {\bibfnamefont {A.~N.}\ \bibnamefont {Tharrington}},\
  }\bibfield  {title} {\enquote {\bibinfo {title} {Implementing molecular
  dynamics on hybrid high performance computers \textendash{} short range
  forces},}\ }\href {\doibase 10.1016/j.cpc.2010.12.021} {\bibfield  {journal}
  {\bibinfo  {journal} {Comput. Phys. Commun.}\ }\textbf {\bibinfo {volume}
  {182}},\ \bibinfo {pages} {898--911} (\bibinfo {year} {2011})}\BibitemShut
  {NoStop}%
\bibitem [{\citenamefont {Brown}\ \emph {et~al.}(2012)\citenamefont {Brown},
  \citenamefont {Kohlmeyer}, \citenamefont {Plimpton},\ and\ \citenamefont
  {Tharrington}}]{Brown2012}%
  \BibitemOpen
  \bibfield  {author} {\bibinfo {author} {\bibfnamefont {W.~M.}\ \bibnamefont
  {Brown}}, \bibinfo {author} {\bibfnamefont {A.}~\bibnamefont {Kohlmeyer}},
  \bibinfo {author} {\bibfnamefont {S.~J.}\ \bibnamefont {Plimpton}}, \ and\
  \bibinfo {author} {\bibfnamefont {A.~N.}\ \bibnamefont {Tharrington}},\
  }\bibfield  {title} {\enquote {\bibinfo {title} {Implementing molecular
  dynamics on hybrid high performance computers \textendash{}
  particle\textendash particle particle\textendash mesh},}\ }\href {\doibase
  10.1016/j.cpc.2011.10.012} {\bibfield  {journal} {\bibinfo  {journal}
  {Comput. Phys. Commun.}\ }\textbf {\bibinfo {volume} {183}},\ \bibinfo
  {pages} {449--459} (\bibinfo {year} {2012})}\BibitemShut {NoStop}%
\bibitem [{\citenamefont {Jorgensen}, \citenamefont {Maxwell},\ and\
  \citenamefont {{Tirado-Rives}}(1996)}]{Jorgensen1996}%
  \BibitemOpen
  \bibfield  {author} {\bibinfo {author} {\bibfnamefont {W.~L.}\ \bibnamefont
  {Jorgensen}}, \bibinfo {author} {\bibfnamefont {D.~S.}\ \bibnamefont
  {Maxwell}}, \ and\ \bibinfo {author} {\bibfnamefont {J.}~\bibnamefont
  {{Tirado-Rives}}},\ }\bibfield  {title} {\enquote {\bibinfo {title}
  {Development and testing of the {{OPLS}} all-atom force field on
  conformational energetics and properties of organic liquids},}\ }\href@noop
  {} {\bibfield  {journal} {\bibinfo  {journal} {J. Am. Chem. Soc.}\ }\textbf
  {\bibinfo {volume} {118}},\ \bibinfo {pages} {11225--11236} (\bibinfo {year}
  {1996})}\BibitemShut {NoStop}%
\bibitem [{\citenamefont {Jorgensen}\ and\ \citenamefont
  {McDonald}(1998)}]{Jorgensen1998}%
  \BibitemOpen
  \bibfield  {author} {\bibinfo {author} {\bibfnamefont {W.~L.}\ \bibnamefont
  {Jorgensen}}\ and\ \bibinfo {author} {\bibfnamefont {N.~A.}\ \bibnamefont
  {McDonald}},\ }\bibfield  {title} {\enquote {\bibinfo {title} {Development of
  an all-atom force field for heterocycles. {{Properties}} of liquid pyridine
  and diazenes},}\ }\href {\doibase 10.1016/S0166-1280(97)00237-6} {\bibfield
  {journal} {\bibinfo  {journal} {J. Mol. Struct. THEOCHEM}\ }\textbf {\bibinfo
  {volume} {424}},\ \bibinfo {pages} {145--155} (\bibinfo {year}
  {1998})}\BibitemShut {NoStop}%
\bibitem [{\citenamefont {Rizzo}\ and\ \citenamefont
  {Jorgensen}(1999)}]{rizzo1999}%
  \BibitemOpen
  \bibfield  {author} {\bibinfo {author} {\bibfnamefont {R.~C.}\ \bibnamefont
  {Rizzo}}\ and\ \bibinfo {author} {\bibfnamefont {W.~L.}\ \bibnamefont
  {Jorgensen}},\ }\bibfield  {title} {\enquote {\bibinfo {title} {{{OPLS}}
  all-atom model for amines: {{Resolution}} of the amine hydration problem},}\
  }\href {\doibase 10.1021/ja984106u} {\bibfield  {journal} {\bibinfo
  {journal} {J. Am. Chem. Soc.}\ }\textbf {\bibinfo {volume} {121}},\ \bibinfo
  {pages} {4827--4836} (\bibinfo {year} {1999})}\BibitemShut {NoStop}%
\bibitem [{\citenamefont {Price}, \citenamefont {Ostrovsky},\ and\
  \citenamefont {Jorgensen}(2001)}]{Price2001}%
  \BibitemOpen
  \bibfield  {author} {\bibinfo {author} {\bibfnamefont {M.~L.}\ \bibnamefont
  {Price}}, \bibinfo {author} {\bibfnamefont {D.}~\bibnamefont {Ostrovsky}}, \
  and\ \bibinfo {author} {\bibfnamefont {W.~L.}\ \bibnamefont {Jorgensen}},\
  }\bibfield  {title} {\enquote {\bibinfo {title} {Gas-phase and liquid-state
  properties of esters, nitriles, and nitro compounds with the {{OPLS-AA}}
  force field},}\ }\href {\doibase 10.1002/JCC.1092} {\bibfield  {journal}
  {\bibinfo  {journal} {J. Comput. Chem.}\ }\textbf {\bibinfo {volume} {22}},\
  \bibinfo {pages} {1340--1352} (\bibinfo {year} {2001})}\BibitemShut {NoStop}%
\bibitem [{\citenamefont {Hockney}\ and\ \citenamefont
  {Eastwood}(1998)}]{Hockney1998}%
  \BibitemOpen
  \bibfield  {author} {\bibinfo {author} {\bibfnamefont {R.}~\bibnamefont
  {Hockney}}\ and\ \bibinfo {author} {\bibfnamefont {J.}~\bibnamefont
  {Eastwood}},\ }\href@noop {} {\emph {\bibinfo {title} {Computer Simulation
  Using Particles}}}\ (\bibinfo  {publisher} {CRC Press},\ \bibinfo {year}
  {1998})\BibitemShut {NoStop}%
\bibitem [{\citenamefont {Ryckaert}, \citenamefont {Ciccotti},\ and\
  \citenamefont {Berendsen}(1977)}]{ryckaert1977numerical}%
  \BibitemOpen
  \bibfield  {author} {\bibinfo {author} {\bibfnamefont {J.-P.}\ \bibnamefont
  {Ryckaert}}, \bibinfo {author} {\bibfnamefont {G.}~\bibnamefont {Ciccotti}},
  \ and\ \bibinfo {author} {\bibfnamefont {H.~J.}\ \bibnamefont {Berendsen}},\
  }\bibfield  {title} {\enquote {\bibinfo {title} {Numerical integration of the
  cartesian equations of motion of a system with constraints: molecular
  dynamics of n-alkanes},}\ }\href@noop {} {\bibfield  {journal} {\bibinfo
  {journal} {J. Comput. Phys.}\ }\textbf {\bibinfo {volume} {23}},\ \bibinfo
  {pages} {327--341} (\bibinfo {year} {1977})}\BibitemShut {NoStop}%
\bibitem [{\citenamefont {Hoover}(1985)}]{hoover1985canonical}%
  \BibitemOpen
  \bibfield  {author} {\bibinfo {author} {\bibfnamefont {W.~G.}\ \bibnamefont
  {Hoover}},\ }\bibfield  {title} {\enquote {\bibinfo {title} {Canonical
  dynamics: Equilibrium phase-space distributions},}\ }\href@noop {} {\bibfield
   {journal} {\bibinfo  {journal} {Phys. Rev. A}\ }\textbf {\bibinfo {volume}
  {31}},\ \bibinfo {pages} {1695} (\bibinfo {year} {1985})}\BibitemShut
  {NoStop}%
\bibitem [{\citenamefont {Nos{\'e}}(1984)}]{nose1984molecular}%
  \BibitemOpen
  \bibfield  {author} {\bibinfo {author} {\bibfnamefont {S.}~\bibnamefont
  {Nos{\'e}}},\ }\bibfield  {title} {\enquote {\bibinfo {title} {A molecular
  dynamics method for simulations in the canonical ensemble},}\ }\href@noop {}
  {\bibfield  {journal} {\bibinfo  {journal} {Mol. Phys.}\ }\textbf {\bibinfo
  {volume} {52}},\ \bibinfo {pages} {255--268} (\bibinfo {year}
  {1984})}\BibitemShut {NoStop}%
\bibitem [{\citenamefont {Marcot}\ and\ \citenamefont
  {Hanea}(2021)}]{Marcot2021}%
  \BibitemOpen
  \bibfield  {author} {\bibinfo {author} {\bibfnamefont {B.~G.}\ \bibnamefont
  {Marcot}}\ and\ \bibinfo {author} {\bibfnamefont {A.~M.}\ \bibnamefont
  {Hanea}},\ }\bibfield  {title} {\enquote {\bibinfo {title} {What is an
  optimal value of k in k-fold cross-validation in discrete {{Bayesian}}
  network analysis?}}\ }\href@noop {} {\bibfield  {journal} {\bibinfo
  {journal} {Comput. Statist.}\ }\textbf {\bibinfo {volume} {36}},\ \bibinfo
  {pages} {2009--2031} (\bibinfo {year} {2021})}\BibitemShut {NoStop}%
\bibitem [{\citenamefont {Bengio}\ and\ \citenamefont
  {Grandvalet}(2003)}]{Bengio2003}%
  \BibitemOpen
  \bibfield  {author} {\bibinfo {author} {\bibfnamefont {Y.}~\bibnamefont
  {Bengio}}\ and\ \bibinfo {author} {\bibfnamefont {Y.}~\bibnamefont
  {Grandvalet}},\ }\bibfield  {title} {\enquote {\bibinfo {title} {No unbiased
  estimator of the variance of k-fold cross-validation},}\ }\href@noop {}
  {\bibfield  {journal} {\bibinfo  {journal} {Adv. Neural Inf. Process. Syst.}\
  }\textbf {\bibinfo {volume} {16}} (\bibinfo {year} {2003})}\BibitemShut
  {NoStop}%
\bibitem [{\citenamefont {Katz}(1997)}]{katz1997organic}%
  \BibitemOpen
  \bibfield  {author} {\bibinfo {author} {\bibfnamefont {H.}~\bibnamefont
  {Katz}},\ }\bibfield  {title} {\enquote {\bibinfo {title} {Organic molecular
  solids as thin film transistor semiconductors},}\ }\href@noop {} {\bibfield
  {journal} {\bibinfo  {journal} {J. Mater. Chem.}\ }\textbf {\bibinfo {volume}
  {7}},\ \bibinfo {pages} {369--376} (\bibinfo {year} {1997})}\BibitemShut
  {NoStop}%
\bibitem [{\citenamefont {Fichou}(2000)}]{Fichou2000}%
  \BibitemOpen
  \bibfield  {author} {\bibinfo {author} {\bibfnamefont {D.}~\bibnamefont
  {Fichou}},\ }\bibfield  {title} {\enquote {\bibinfo {title} {Structural order
  in conjugated oligothiophenes and its implications on opto-electronic
  devices},}\ }\href {\doibase 10.1039/A908312J} {\bibfield  {journal}
  {\bibinfo  {journal} {J. Mater. Chem.}\ }\textbf {\bibinfo {volume} {10}},\
  \bibinfo {pages} {571--588} (\bibinfo {year} {2000})}\BibitemShut {NoStop}%
\bibitem [{\citenamefont {Dong}\ \emph {et~al.}(2020)\citenamefont {Dong},
  \citenamefont {Nikolis}, \citenamefont {Talnack}, \citenamefont {Chin},
  \citenamefont {Benduhn}, \citenamefont {Londi}, \citenamefont {Kublitski},
  \citenamefont {Zheng}, \citenamefont {Mannsfeld}, \citenamefont {Spoltore},
  \citenamefont {Muccioli}, \citenamefont {Li}, \citenamefont {Blase},
  \citenamefont {Beljonne}, \citenamefont {Kim}, \citenamefont {Bakulin},
  \citenamefont {D'Avino}, \citenamefont {Durrant},\ and\ \citenamefont
  {Vandewal}}]{dong2020orientation}%
  \BibitemOpen
  \bibfield  {author} {\bibinfo {author} {\bibfnamefont {Y.}~\bibnamefont
  {Dong}}, \bibinfo {author} {\bibfnamefont {V.~C.}\ \bibnamefont {Nikolis}},
  \bibinfo {author} {\bibfnamefont {F.}~\bibnamefont {Talnack}}, \bibinfo
  {author} {\bibfnamefont {Y.-C.}\ \bibnamefont {Chin}}, \bibinfo {author}
  {\bibfnamefont {J.}~\bibnamefont {Benduhn}}, \bibinfo {author} {\bibfnamefont
  {G.}~\bibnamefont {Londi}}, \bibinfo {author} {\bibfnamefont
  {J.}~\bibnamefont {Kublitski}}, \bibinfo {author} {\bibfnamefont
  {X.}~\bibnamefont {Zheng}}, \bibinfo {author} {\bibfnamefont {S.~C.~B.}\
  \bibnamefont {Mannsfeld}}, \bibinfo {author} {\bibfnamefont {D.}~\bibnamefont
  {Spoltore}}, \bibinfo {author} {\bibfnamefont {L.}~\bibnamefont {Muccioli}},
  \bibinfo {author} {\bibfnamefont {J.}~\bibnamefont {Li}}, \bibinfo {author}
  {\bibfnamefont {X.}~\bibnamefont {Blase}}, \bibinfo {author} {\bibfnamefont
  {D.}~\bibnamefont {Beljonne}}, \bibinfo {author} {\bibfnamefont {J.-S.}\
  \bibnamefont {Kim}}, \bibinfo {author} {\bibfnamefont {A.~A.}\ \bibnamefont
  {Bakulin}}, \bibinfo {author} {\bibfnamefont {G.}~\bibnamefont {D'Avino}},
  \bibinfo {author} {\bibfnamefont {J.~R.}\ \bibnamefont {Durrant}}, \ and\
  \bibinfo {author} {\bibfnamefont {K.}~\bibnamefont {Vandewal}},\ }\bibfield
  {title} {\enquote {\bibinfo {title} {Orientation dependent molecular
  electrostatics drives efficient charge generation in homojunction organic
  solar cells},}\ }\href@noop {} {\bibfield  {journal} {\bibinfo  {journal}
  {Nat. Commun.}\ }\textbf {\bibinfo {volume} {11}},\ \bibinfo {pages} {4617}
  (\bibinfo {year} {2020})}\BibitemShut {NoStop}%
\bibitem [{\citenamefont {Haley}\ and\ \citenamefont
  {Soloway}(1992)}]{Haley1992}%
  \BibitemOpen
  \bibfield  {author} {\bibinfo {author} {\bibfnamefont {P.~J.}\ \bibnamefont
  {Haley}}\ and\ \bibinfo {author} {\bibfnamefont {D.}~\bibnamefont
  {Soloway}},\ }\bibfield  {title} {\enquote {\bibinfo {title} {Extrapolation
  limitations of multilayer feedforward neural networks},}\ }in\ \href@noop {}
  {\emph {\bibinfo {booktitle} {Proc. 1992 {{IJCNN Int}}. {{Jt}}. {{Conf}}.
  {{Neural Netw}}.}}},\ Vol.~\bibinfo {volume} {4}\ (\bibinfo {organization}
  {{IEEE}},\ \bibinfo {year} {1992})\ pp.\ \bibinfo {pages}
  {25--30}\BibitemShut {NoStop}%
\bibitem [{\citenamefont {Na}, \citenamefont {Jang},\ and\ \citenamefont
  {Chang}(2022)}]{Na2022}%
  \BibitemOpen
  \bibfield  {author} {\bibinfo {author} {\bibfnamefont {G.~S.}\ \bibnamefont
  {Na}}, \bibinfo {author} {\bibfnamefont {S.}~\bibnamefont {Jang}}, \ and\
  \bibinfo {author} {\bibfnamefont {H.}~\bibnamefont {Chang}},\ }\bibfield
  {title} {\enquote {\bibinfo {title} {Nonlinearity encoding to improve
  extrapolation capabilities for unobserved physical states},}\ }\href@noop {}
  {\bibfield  {journal} {\bibinfo  {journal} {Phys. Chem. Chem. Phys.}\
  }\textbf {\bibinfo {volume} {24}},\ \bibinfo {pages} {1300--1304} (\bibinfo
  {year} {2022})}\BibitemShut {NoStop}%
\bibitem [{\citenamefont {Ding}\ \emph {et~al.}(2021)\citenamefont {Ding},
  \citenamefont {Pervaiz}, \citenamefont {Carbin},\ and\ \citenamefont
  {Hoffmann}}]{Ding2021}%
  \BibitemOpen
  \bibfield  {author} {\bibinfo {author} {\bibfnamefont {Y.}~\bibnamefont
  {Ding}}, \bibinfo {author} {\bibfnamefont {A.}~\bibnamefont {Pervaiz}},
  \bibinfo {author} {\bibfnamefont {M.}~\bibnamefont {Carbin}}, \ and\ \bibinfo
  {author} {\bibfnamefont {H.}~\bibnamefont {Hoffmann}},\ }\bibfield  {title}
  {\enquote {\bibinfo {title} {Generalizable and interpretable learning for
  configuration extrapolation},}\ }in\ \href@noop {} {\emph {\bibinfo
  {booktitle} {Proc. 29th {{ACM Jt}}. {{Meet}}. {{Eur}}. {{Softw}}. {{Eng}}.
  {{Conf}}. {{Symp}}. {{Found}}. {{Softw}}. {{Eng}}.}}}\ (\bibinfo {year}
  {2021})\ pp.\ \bibinfo {pages} {728--740}\BibitemShut {NoStop}%
\bibitem [{\citenamefont {Rigol}, \citenamefont {Jarvis},\ and\ \citenamefont
  {Stuart}(2001)}]{Rigol2001}%
  \BibitemOpen
  \bibfield  {author} {\bibinfo {author} {\bibfnamefont {J.~P.}\ \bibnamefont
  {Rigol}}, \bibinfo {author} {\bibfnamefont {C.~H.}\ \bibnamefont {Jarvis}}, \
  and\ \bibinfo {author} {\bibfnamefont {N.}~\bibnamefont {Stuart}},\
  }\bibfield  {title} {\enquote {\bibinfo {title} {Artificial neural networks
  as a tool for spatial interpolation},}\ }\href@noop {} {\bibfield  {journal}
  {\bibinfo  {journal} {Int. J. Geogr. Inf. Sci.}\ }\textbf {\bibinfo {volume}
  {15}},\ \bibinfo {pages} {323--343} (\bibinfo {year} {2001})}\BibitemShut
  {NoStop}%
\bibitem [{\citenamefont {Mueller}\ \emph {et~al.}(2015)\citenamefont
  {Mueller}, \citenamefont {Zierenberg}, \citenamefont {Marenz}, \citenamefont
  {Schierz},\ and\ \citenamefont {Janke}}]{Mueller2015}%
  \BibitemOpen
  \bibfield  {author} {\bibinfo {author} {\bibfnamefont {M.}~\bibnamefont
  {Mueller}}, \bibinfo {author} {\bibfnamefont {J.}~\bibnamefont {Zierenberg}},
  \bibinfo {author} {\bibfnamefont {M.}~\bibnamefont {Marenz}}, \bibinfo
  {author} {\bibfnamefont {P.}~\bibnamefont {Schierz}}, \ and\ \bibinfo
  {author} {\bibfnamefont {W.}~\bibnamefont {Janke}},\ }\bibfield  {title}
  {\enquote {\bibinfo {title} {Probing the effect of density on the aggregation
  temperature of semi-flexible polymers in spherical confinement},}\
  }\href@noop {} {\bibfield  {journal} {\bibinfo  {journal} {Phys. Procedia}\
  }\textbf {\bibinfo {volume} {68}},\ \bibinfo {pages} {95--99} (\bibinfo
  {year} {2015})}\BibitemShut {NoStop}%
\bibitem [{\citenamefont {Seaton}, \citenamefont {Mitchell},\ and\
  \citenamefont {Landau}(2006)}]{Seaton2006}%
  \BibitemOpen
  \bibfield  {author} {\bibinfo {author} {\bibfnamefont {D.}~\bibnamefont
  {Seaton}}, \bibinfo {author} {\bibfnamefont {S.}~\bibnamefont {Mitchell}}, \
  and\ \bibinfo {author} {\bibfnamefont {D.}~\bibnamefont {Landau}},\
  }\bibfield  {title} {\enquote {\bibinfo {title} {Monte {{Carlo}} simulations
  of a semi-flexible polymer chain: a first glance},}\ }\href@noop {}
  {\bibfield  {journal} {\bibinfo  {journal} {Braz. J. Phys.}\ }\textbf
  {\bibinfo {volume} {36}},\ \bibinfo {pages} {623--626} (\bibinfo {year}
  {2006})}\BibitemShut {NoStop}%
\bibitem [{\citenamefont {Sinitskiy}, \citenamefont {Saunders},\ and\
  \citenamefont {Voth}(2012)}]{Sinitskiy2012}%
  \BibitemOpen
  \bibfield  {author} {\bibinfo {author} {\bibfnamefont {A.~V.}\ \bibnamefont
  {Sinitskiy}}, \bibinfo {author} {\bibfnamefont {M.~G.}\ \bibnamefont
  {Saunders}}, \ and\ \bibinfo {author} {\bibfnamefont {G.~A.}\ \bibnamefont
  {Voth}},\ }\bibfield  {title} {\enquote {\bibinfo {title} {Optimal number of
  coarse-grained sites in different components of large biomolecular
  complexes},}\ }\href {\doibase 10.1021/jp2108895} {\bibfield  {journal}
  {\bibinfo  {journal} {J. Phys. Chem. B}\ }\textbf {\bibinfo {volume} {116}},\
  \bibinfo {pages} {8363--8374} (\bibinfo {year} {2012})}\BibitemShut {NoStop}%
\bibitem [{\citenamefont {Falkowska}\ \emph {et~al.}(2016)\citenamefont
  {Falkowska}, \citenamefont {Bowron}, \citenamefont {Manyar}, \citenamefont
  {Hardacre},\ and\ \citenamefont {Youngs}}]{Falkowska2016}%
  \BibitemOpen
  \bibfield  {author} {\bibinfo {author} {\bibfnamefont {M.}~\bibnamefont
  {Falkowska}}, \bibinfo {author} {\bibfnamefont {D.~T.}\ \bibnamefont
  {Bowron}}, \bibinfo {author} {\bibfnamefont {H.~G.}\ \bibnamefont {Manyar}},
  \bibinfo {author} {\bibfnamefont {C.}~\bibnamefont {Hardacre}}, \ and\
  \bibinfo {author} {\bibfnamefont {T.~G.}\ \bibnamefont {Youngs}},\ }\bibfield
   {title} {\enquote {\bibinfo {title} {Neutron scattering of aromatic and
  aliphatic liquids},}\ }\href {\doibase 10.1002/cphc.201600149} {\bibfield
  {journal} {\bibinfo  {journal} {ChemPhysChem}\ }\textbf {\bibinfo {volume}
  {17}},\ \bibinfo {pages} {2043--2055} (\bibinfo {year} {2016})}\BibitemShut
  {NoStop}%
\bibitem [{\citenamefont {Pizzirusso}\ \emph {et~al.}(2011)\citenamefont
  {Pizzirusso}, \citenamefont {Savini}, \citenamefont {Muccioli},\ and\
  \citenamefont {Zannoni}}]{Pizzirusso2011}%
  \BibitemOpen
  \bibfield  {author} {\bibinfo {author} {\bibfnamefont {A.}~\bibnamefont
  {Pizzirusso}}, \bibinfo {author} {\bibfnamefont {M.}~\bibnamefont {Savini}},
  \bibinfo {author} {\bibfnamefont {L.}~\bibnamefont {Muccioli}}, \ and\
  \bibinfo {author} {\bibfnamefont {C.}~\bibnamefont {Zannoni}},\ }\bibfield
  {title} {\enquote {\bibinfo {title} {An atomistic simulation of the
  liquid-crystalline phases of sexithiophene},}\ }\href {\doibase
  10.1039/C0JM01284J} {\bibfield  {journal} {\bibinfo  {journal} {J. Mater.
  Chem.}\ }\textbf {\bibinfo {volume} {21}},\ \bibinfo {pages} {125--133}
  (\bibinfo {year} {2011})}\BibitemShut {NoStop}%
\bibitem [{\citenamefont {Weiner}\ \emph {et~al.}(1984)\citenamefont {Weiner},
  \citenamefont {Kollman}, \citenamefont {Case}, \citenamefont {Singh},
  \citenamefont {Ghio}, \citenamefont {Alagona}, \citenamefont {Profeta},\ and\
  \citenamefont {Weiner}}]{weiner1984new}%
  \BibitemOpen
  \bibfield  {author} {\bibinfo {author} {\bibfnamefont {S.~J.}\ \bibnamefont
  {Weiner}}, \bibinfo {author} {\bibfnamefont {P.~A.}\ \bibnamefont {Kollman}},
  \bibinfo {author} {\bibfnamefont {D.~A.}\ \bibnamefont {Case}}, \bibinfo
  {author} {\bibfnamefont {U.~C.}\ \bibnamefont {Singh}}, \bibinfo {author}
  {\bibfnamefont {C.}~\bibnamefont {Ghio}}, \bibinfo {author} {\bibfnamefont
  {G.}~\bibnamefont {Alagona}}, \bibinfo {author} {\bibfnamefont
  {S.}~\bibnamefont {Profeta}}, \ and\ \bibinfo {author} {\bibfnamefont
  {P.}~\bibnamefont {Weiner}},\ }\bibfield  {title} {\enquote {\bibinfo {title}
  {A new force field for molecular mechanical simulation of nucleic acids and
  proteins},}\ }\href@noop {} {\bibfield  {journal} {\bibinfo  {journal} {J.
  Am. Chem. Soc.}\ }\textbf {\bibinfo {volume} {106}},\ \bibinfo {pages}
  {765--784} (\bibinfo {year} {1984})}\BibitemShut {NoStop}%
\bibitem [{\citenamefont {Weiner}\ \emph {et~al.}(1986)\citenamefont {Weiner},
  \citenamefont {Kollman}, \citenamefont {Nguyen},\ and\ \citenamefont
  {Case}}]{weiner1986all}%
  \BibitemOpen
  \bibfield  {author} {\bibinfo {author} {\bibfnamefont {S.~J.}\ \bibnamefont
  {Weiner}}, \bibinfo {author} {\bibfnamefont {P.~A.}\ \bibnamefont {Kollman}},
  \bibinfo {author} {\bibfnamefont {D.~T.}\ \bibnamefont {Nguyen}}, \ and\
  \bibinfo {author} {\bibfnamefont {D.~A.}\ \bibnamefont {Case}},\ }\bibfield
  {title} {\enquote {\bibinfo {title} {An all atom force field for simulations
  of proteins and nucleic acids},}\ }\href@noop {} {\bibfield  {journal}
  {\bibinfo  {journal} {J. Comput. Chem.}\ }\textbf {\bibinfo {volume} {7}},\
  \bibinfo {pages} {230--252} (\bibinfo {year} {1986})}\BibitemShut {NoStop}%
\bibitem [{\citenamefont {Cornell}\ \emph {et~al.}(1995)\citenamefont
  {Cornell}, \citenamefont {Cieplak}, \citenamefont {Bayly}, \citenamefont
  {Gould}, \citenamefont {Merz}, \citenamefont {Ferguson}, \citenamefont
  {Spellmeyer}, \citenamefont {Fox}, \citenamefont {Caldwell},\ and\
  \citenamefont {Kollman}}]{cornell1995second}%
  \BibitemOpen
  \bibfield  {author} {\bibinfo {author} {\bibfnamefont {W.~D.}\ \bibnamefont
  {Cornell}}, \bibinfo {author} {\bibfnamefont {P.}~\bibnamefont {Cieplak}},
  \bibinfo {author} {\bibfnamefont {C.~I.}\ \bibnamefont {Bayly}}, \bibinfo
  {author} {\bibfnamefont {I.~R.}\ \bibnamefont {Gould}}, \bibinfo {author}
  {\bibfnamefont {K.~M.}\ \bibnamefont {Merz}}, \bibinfo {author}
  {\bibfnamefont {D.~M.}\ \bibnamefont {Ferguson}}, \bibinfo {author}
  {\bibfnamefont {D.~C.}\ \bibnamefont {Spellmeyer}}, \bibinfo {author}
  {\bibfnamefont {T.}~\bibnamefont {Fox}}, \bibinfo {author} {\bibfnamefont
  {J.~W.}\ \bibnamefont {Caldwell}}, \ and\ \bibinfo {author} {\bibfnamefont
  {P.~A.}\ \bibnamefont {Kollman}},\ }\bibfield  {title} {\enquote {\bibinfo
  {title} {A second generation force field for the simulation of proteins,
  nucleic acids, and organic molecules},}\ }\href@noop {} {\bibfield  {journal}
  {\bibinfo  {journal} {J. Am. Chem. Soc.}\ }\textbf {\bibinfo {volume}
  {117}},\ \bibinfo {pages} {5179--5197} (\bibinfo {year} {1995})}\BibitemShut
  {NoStop}%
\bibitem [{\citenamefont {Tsourtou}\ \emph {et~al.}(2018)\citenamefont
  {Tsourtou}, \citenamefont {Peroukidis}, \citenamefont {Peristeras},\ and\
  \citenamefont {Mavrantzas}}]{Tsourtou2018a}%
  \BibitemOpen
  \bibfield  {author} {\bibinfo {author} {\bibfnamefont {F.~D.}\ \bibnamefont
  {Tsourtou}}, \bibinfo {author} {\bibfnamefont {S.~D.}\ \bibnamefont
  {Peroukidis}}, \bibinfo {author} {\bibfnamefont {L.~D.}\ \bibnamefont
  {Peristeras}}, \ and\ \bibinfo {author} {\bibfnamefont {V.~G.}\ \bibnamefont
  {Mavrantzas}},\ }\bibfield  {title} {\enquote {\bibinfo {title} {Monte
  {{Carlo}} algorithm based on internal bridging moves for the atomistic
  simulation of thiophene oligomers and polymers},}\ }\href@noop {} {\bibfield
  {journal} {\bibinfo  {journal} {Macromolecules}\ }\textbf {\bibinfo {volume}
  {51}},\ \bibinfo {pages} {8406--8423} (\bibinfo {year} {2018})}\BibitemShut
  {NoStop}%
\bibitem [{\citenamefont {Xia}\ \emph {et~al.}(2019)\citenamefont {Xia},
  \citenamefont {Hansoge}, \citenamefont {Xu}, \citenamefont {Phelan~Jr},
  \citenamefont {Keten},\ and\ \citenamefont {Douglas}}]{Xia2019}%
  \BibitemOpen
  \bibfield  {author} {\bibinfo {author} {\bibfnamefont {W.}~\bibnamefont
  {Xia}}, \bibinfo {author} {\bibfnamefont {N.~K.}\ \bibnamefont {Hansoge}},
  \bibinfo {author} {\bibfnamefont {W.-S.}\ \bibnamefont {Xu}}, \bibinfo
  {author} {\bibfnamefont {F.~R.}\ \bibnamefont {Phelan~Jr}}, \bibinfo {author}
  {\bibfnamefont {S.}~\bibnamefont {Keten}}, \ and\ \bibinfo {author}
  {\bibfnamefont {J.~F.}\ \bibnamefont {Douglas}},\ }\bibfield  {title}
  {\enquote {\bibinfo {title} {Energy renormalization for coarse-graining
  polymers having different segmental structures},}\ }\href@noop {} {\bibfield
  {journal} {\bibinfo  {journal} {Sci. Adv.}\ }\textbf {\bibinfo {volume}
  {5}},\ \bibinfo {pages} {eaav4683} (\bibinfo {year} {2019})}\BibitemShut
  {NoStop}%
\bibitem [{\citenamefont {D'Adamo}\ \emph {et~al.}(2015)\citenamefont
  {D'Adamo}, \citenamefont {Menichetti}, \citenamefont {Pelissetto},\ and\
  \citenamefont {Pierleoni}}]{DAdamo2015}%
  \BibitemOpen
  \bibfield  {author} {\bibinfo {author} {\bibfnamefont {G.}~\bibnamefont
  {D'Adamo}}, \bibinfo {author} {\bibfnamefont {R.}~\bibnamefont {Menichetti}},
  \bibinfo {author} {\bibfnamefont {A.}~\bibnamefont {Pelissetto}}, \ and\
  \bibinfo {author} {\bibfnamefont {C.}~\bibnamefont {Pierleoni}},\ }\bibfield
  {title} {\enquote {\bibinfo {title} {Coarse-graining polymer solutions: {{A}}
  critical appraisal of single-and multi-site models},}\ }\href@noop {}
  {\bibfield  {journal} {\bibinfo  {journal} {Eur. Phys. J. Spec. Top.}\
  }\textbf {\bibinfo {volume} {224}},\ \bibinfo {pages} {2239--2267} (\bibinfo
  {year} {2015})}\BibitemShut {NoStop}%
\end{thebibliography}%


\begin{thebibliography}{1}%
\makeatletter
\providecommand \@ifxundefined [1]{%
 \@ifx{#1\undefined}
}%
\providecommand \@ifnum [1]{%
 \ifnum #1\expandafter \@firstoftwo
 \else \expandafter \@secondoftwo
 \fi
}%
\providecommand \@ifx [1]{%
 \ifx #1\expandafter \@firstoftwo
 \else \expandafter \@secondoftwo
 \fi
}%
\providecommand \natexlab [1]{#1}%
\providecommand \enquote  [1]{``#1''}%
\providecommand \bibnamefont  [1]{#1}%
\providecommand \bibfnamefont [1]{#1}%
\providecommand \citenamefont [1]{#1}%
\providecommand \href@noop [0]{\@secondoftwo}%
\providecommand \href [0]{\begingroup \@sanitize@url \@href}%
\providecommand \@href[1]{\@@startlink{#1}\@@href}%
\providecommand \@@href[1]{\endgroup#1\@@endlink}%
\providecommand \@sanitize@url [0]{\catcode `\\12\catcode `\$12\catcode
  `\&12\catcode `\#12\catcode `\^12\catcode `\_12\catcode `\%12\relax}%
\providecommand \@@startlink[1]{}%
\providecommand \@@endlink[0]{}%
\providecommand \url  [0]{\begingroup\@sanitize@url \@url }%
\providecommand \@url [1]{\endgroup\@href {#1}{\urlprefix }}%
\providecommand \urlprefix  [0]{URL }%
\providecommand \Eprint [0]{\href }%
\providecommand \doibase [0]{http://dx.doi.org/}%
\providecommand \selectlanguage [0]{\@gobble}%
\providecommand \bibinfo  [0]{\@secondoftwo}%
\providecommand \bibfield  [0]{\@secondoftwo}%
\providecommand \translation [1]{[#1]}%
\providecommand \BibitemOpen [0]{}%
\providecommand \bibitemStop [0]{}%
\providecommand \bibitemNoStop [0]{.\EOS\space}%
\providecommand \EOS [0]{\spacefactor3000\relax}%
\providecommand \BibitemShut  [1]{\csname bibitem#1\endcsname}%
\let\auto@bib@innerbib\@empty
\bibitem [{\citenamefont {Kingma}\ and\ \citenamefont {Ba}(2014)}]{Kingma2014}%
  \BibitemOpen
  \bibfield  {author} {\bibinfo {author} {\bibfnamefont {D.~P.}\ \bibnamefont
  {Kingma}}\ and\ \bibinfo {author} {\bibfnamefont {J.}~\bibnamefont {Ba}},\
  }\bibfield  {title} {\enquote {\bibinfo {title} {Adam: {{A}} method for
  stochastic optimization},}\ }\href@noop {} {\bibfield  {journal} {\bibinfo
  {journal} {arXiv}\ ,\ \bibinfo {pages} {1412.6980}} (\bibinfo {year}
  {2014})}\BibitemShut {NoStop}%
\end{thebibliography}%

\end{document}


\begin{center}
	{\large\textbf{Supplementary Material:}} \\
	{\large\textbf{Anisotropic molecular coarse-graining by force and torque matching with neural networks}} \\ 
	Marltan O. Wilson and David M. Huang\\
	\textit{Department of Chemistry, School of Physics, Chemistry and Earth Sciences,\\ The University of Adelaide, Adelaide, South Australia 5005, Australia}
\end{center}
	
	\tableofcontents
	\clearpage

\section{Benzene network parameters}

The cut-off distance $\Rc$ for the benzene neural network was set at 10~\AA\ for the $G^1$ type symmetry function. For the $G^5$ angular symmetry function, $\lambda$ had values of -1 and 1 and $ \nu$ has values of $2^n$, where $n \in \mathbb{Z}$. Hyperparameters $\alpha$, $\beta$, and $\gamma$ in the loss function were adjusted to improve the speed of convergence, but this did not usually affect the global minimum of the optimization when the number of training epochs was large.

\begin{table}[H]
	\caption{\label{tab:table2}Hyperparameters for benzene. }
	\centering
	\begin{tabular}{c | c | c}
		\hline
		hyperparameter	&value &units\\
		\hline
		$\lambda$& {[-1.0, 1.0]} \\
		$\eta$& {[2.0, 1.0]}  &\SI{}{\per\angstrom\squared}\\
		$\nu$& {[2.0, 4.0,  8.0, 16.0, 32.0, 64.0]}  \\
		$\Rs$& {[3.0,3.7, 4.3, 5.0, 5.7, 6.3, 7.0, 7.7, 8.3, 9.0]} &  \SI{}{\angstrom} \\
		$\Rc$& {[10.0]} &  \SI{}{\angstrom}  \\
		\hline
	\end{tabular}
\end{table}

\newpage

\section{Additional benzene structural distributions}

\begin{table}[H]
	\caption{\label{tab:table2}The optimal coarse-grained principal moments of inertia $I_q$ for $q = 1,2,3$, calculated using Eqs.~(10) of the main paper. }
	\centering
	\begin{tabular}{c | c }
	\hline
	 principal axis $q$ & $I_q$ (\si{\gram\per\mole\per\angstrom\squared}) \\
		\hline
		$1$& \num{88.1} \\
		$2$& \num{92.2}\\
		$3$& \num{ 180.1} \\
		\hline
	\end{tabular}
\end{table}

\begin{figure}[H]
	\centering
	\includegraphics{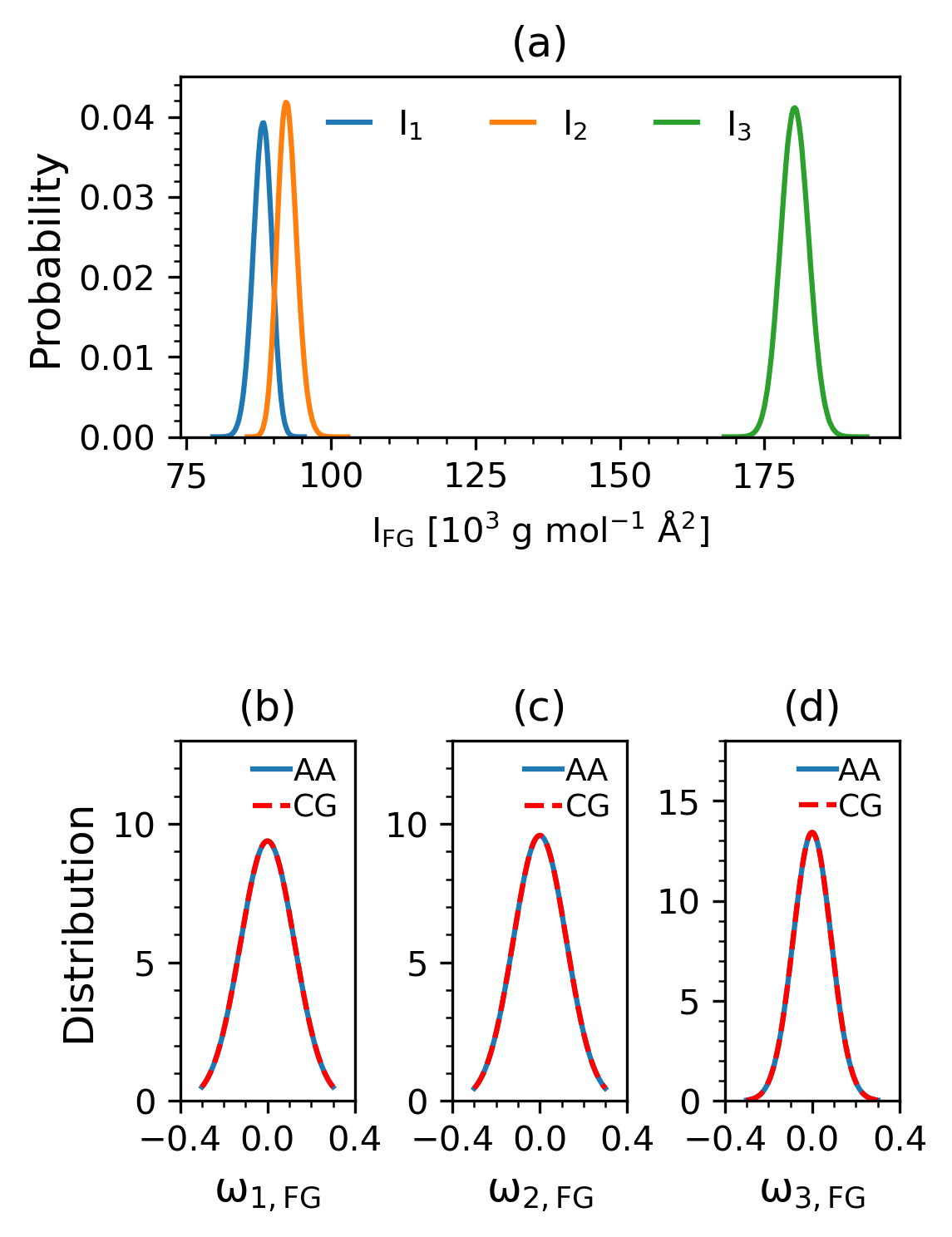}
	\caption{(a) Principal moment of inertia distributions for the all-atom (AA) benzene model at 300~K and 1~atm. The corresponding angular velocity distributions of each principal axis along with the coarse-grained (CG) fit to the distribution given by $I_{q}^{1/2}\exp(-\frac{I_{q}\omega_{q}^2}{2k_BT})$ is shown in (b)--(d).}
	\label{fig:benzeneinertiadistribution2}
\end{figure}

\begin{figure}[H]
	\centering
	\includegraphics{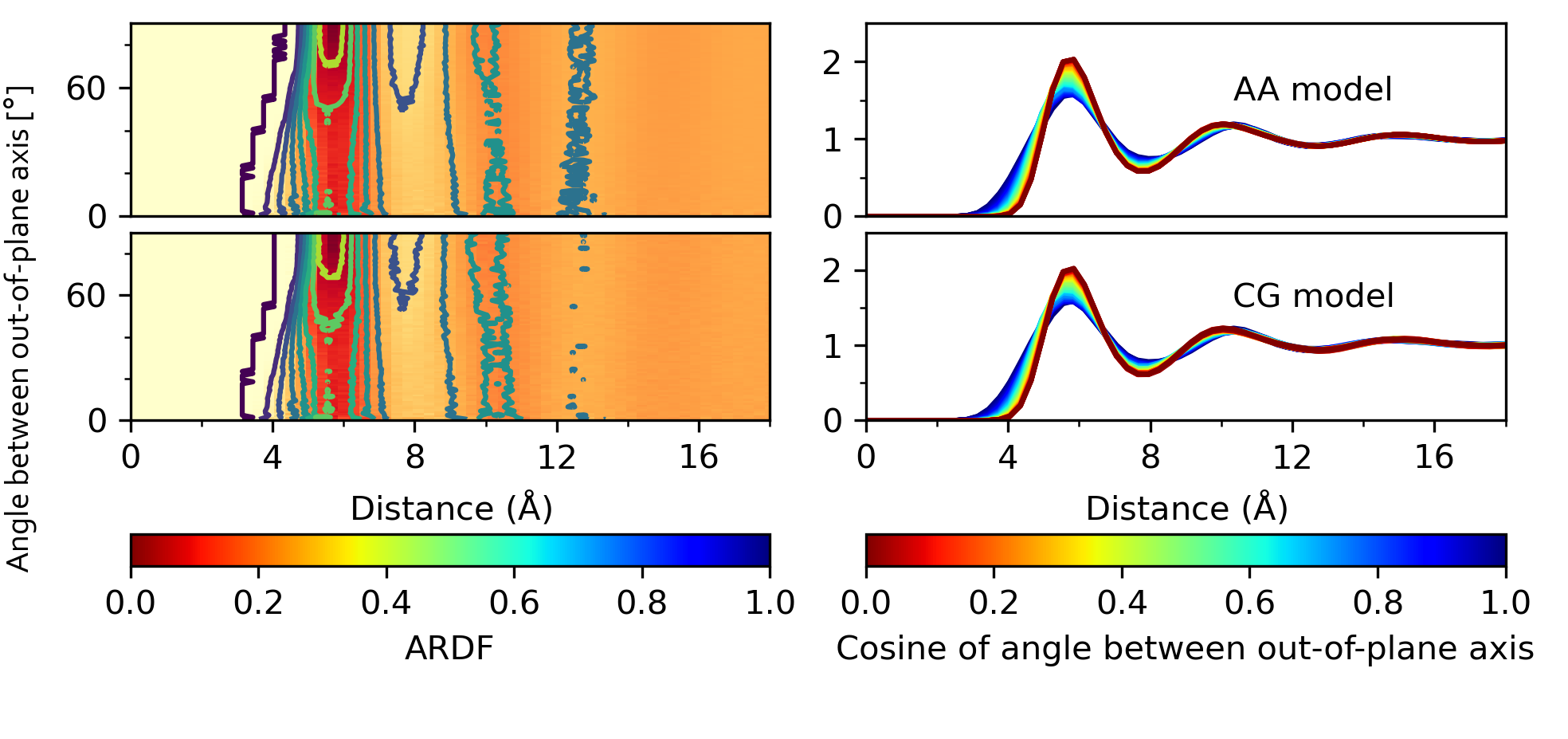}
	\caption{Angular--radial distribution function (ARDF) of the all-atom (AA) (top) and coarse-grained (CG) (bottom) benzene models at 280~K and 1~atm depicted as a heat map (left) and 1D slices at constant angle (right). Face-to-face, edge-to-edge, or parallel displaced configurations occur when the angle is 0\textdegree, while edge-to-face configurations occur at 90\textdegree. }
	\label{fig:benzene280ardf}
\end{figure}

\begin{figure}[H]
	\centering
	\includegraphics{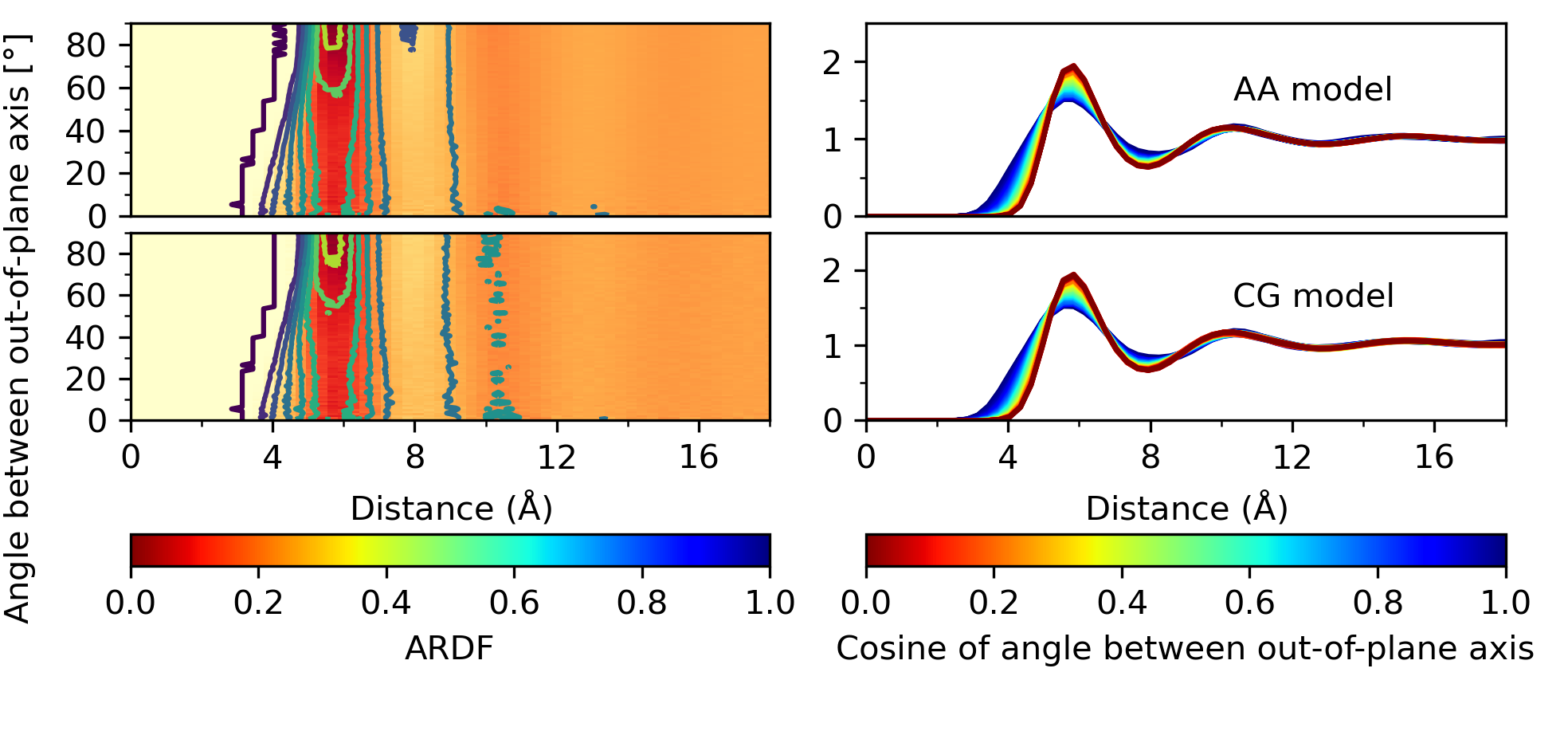}
	\caption{Angular--radial distribution function (ARDF) of the all-atom (AA) (top) and coarse-grained (CG) (bottom) benzene models at 320~K and 1~atm depicted as a heat map (left) and 1D slices at constant angle (right). Face-to-face, edge-to-edge, and parallel displaced configurations occur when the angle is 0\textdegree, while edge-to-face configurations occur at 90\textdegree. }
	\label{fig:benzene320ardf}
\end{figure}

\begin{figure}[H]
	\centering
	\includegraphics{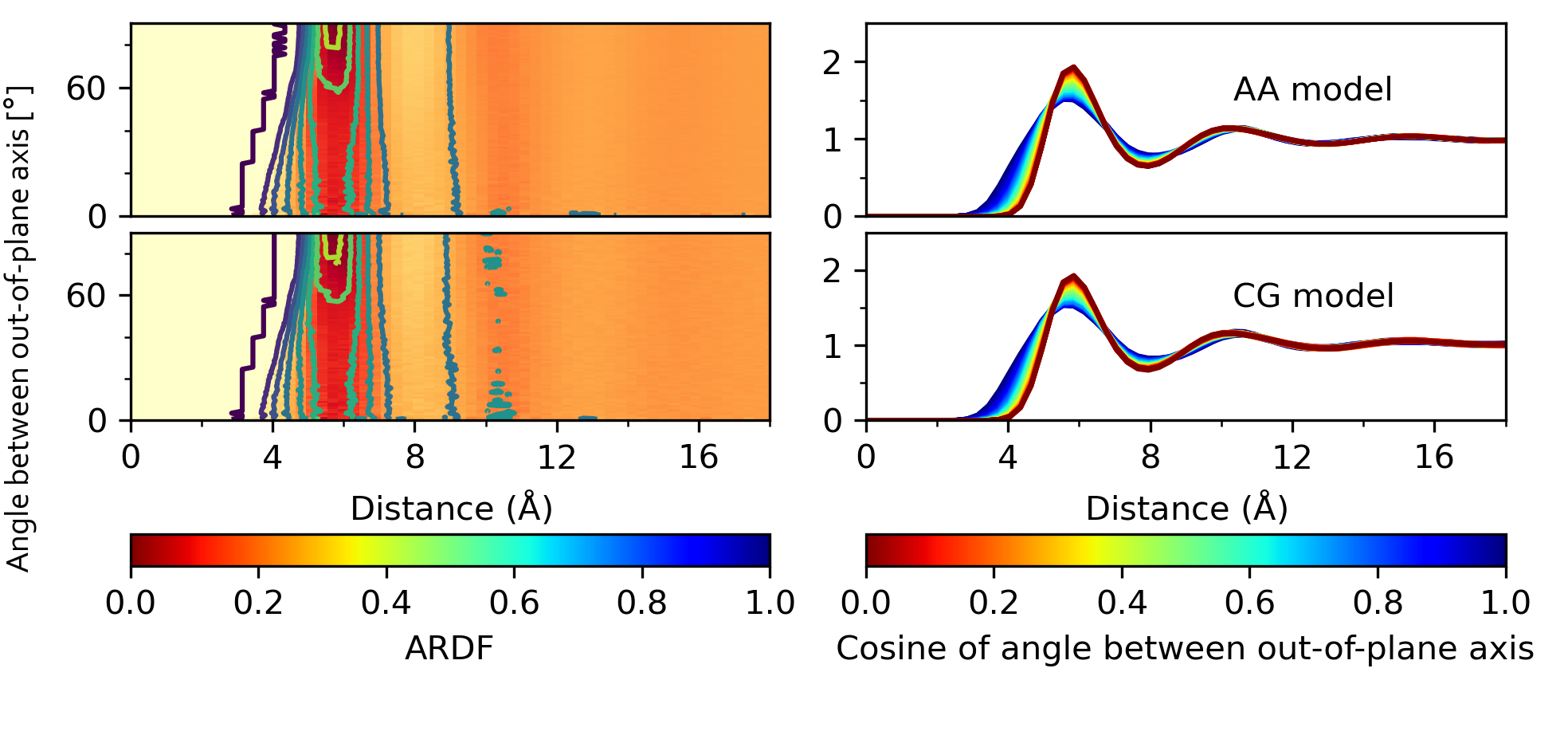}
	\caption{Angular--radial distribution function (ARDF) of the all-atom (AA) (top) and coarse-grained (CG) (bottom) benzene models at 330~K and 1~atm depicted as a heat map (left) and 1D slices at constant angle (right). Face-to-face, edge-to-edge, and parallel displaced configurations occur when the angle is 0\textdegree, while edge-to-face configurations occur at 90\textdegree. }
	\label{fig:benzene330ardf}
\end{figure}

\begin{figure}[H]
	\centering
	\includegraphics{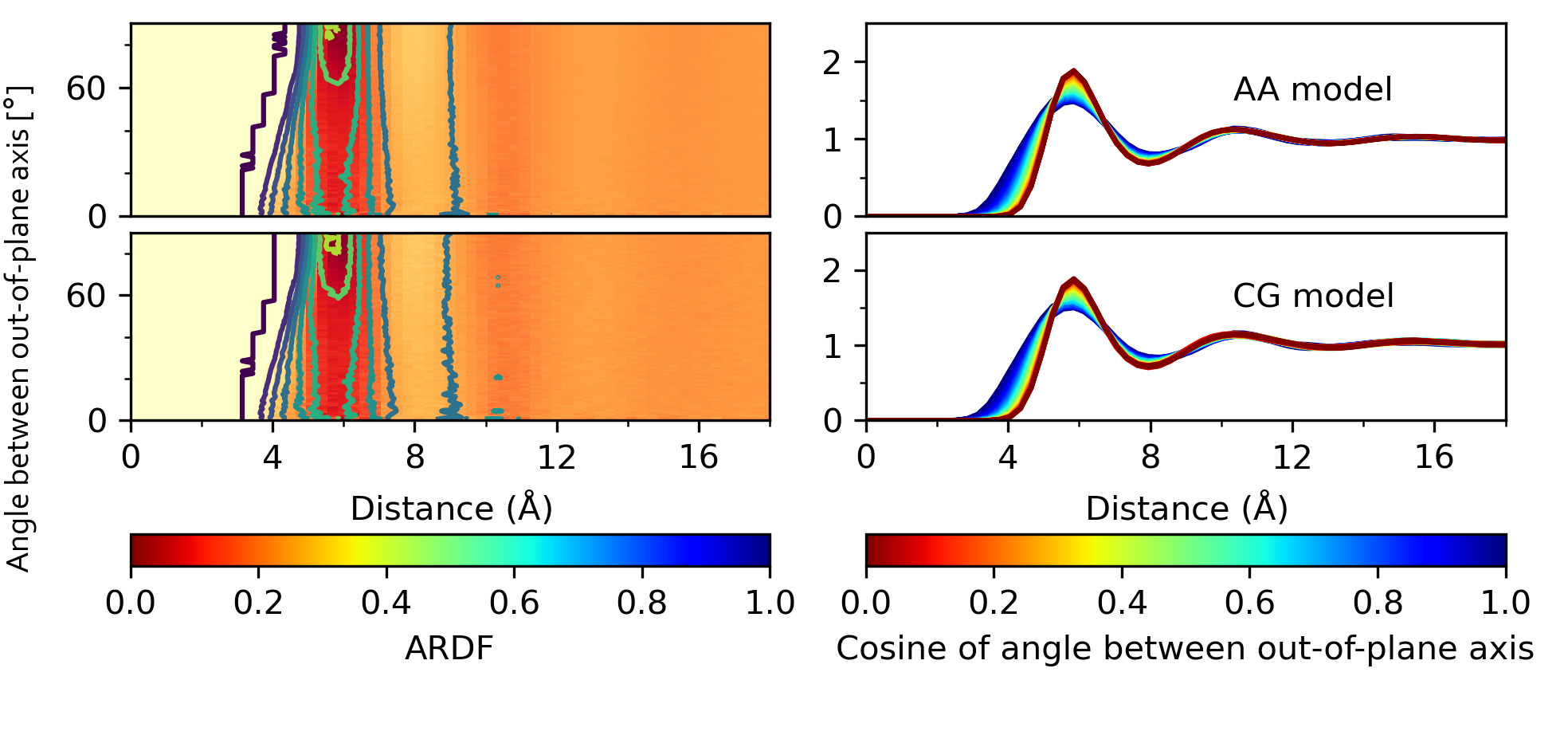}
	\caption{Angular--radial distribution function (ARDF) of the all-atom (AA) (top) and coarse-grained (CG) (bottom) benzene models at 350~K and 1~atm depicted as a heat map (left) and 1D slices at constant angle (right). Face-to-face, edge-to-edge, and parallel displaced configurations occur when the angle is 0\textdegree, while edge-to-face configurations occur at 90\textdegree.}
	\label{fig:benzene350ardf}
\end{figure}

\newpage

\section{Sexithiophene network parameters}

The cut-off distance $\Rc$ for the sexithiophene neural network was set to 21~\AA\ for the $G^1$ type symmetry function. For the $G^5$ angular symmetry function $\lambda$ had values of -1 and 1, $ \nu$ has values of $2^n$ where $n \in \mathbb{Z}$. Hyperparameters $\alpha$, $\beta$, and $\gamma$ in the loss function were adjusted to improve the speed of convergence but did not usually affect the global minimum of the optimization when the number of training epochs was large.

\begin{table}[H]
	\caption{\label{tab:table2}Hyperparameters for sexithiophene}
		\centering
	\begin{tabular}{c | c | c}
	\hline
	 hyperparameter	&value &units\\
		\hline
		$\lambda$& {[-1.0, 1.0]} \\
		$\eta$& {[2.0, 1.0]}  &\SI{}{\per\angstrom\squared}\\
		$\nu$& {[2.0, 4.0,  8.0, 16.0, 32.0, 64.0]}  \\
		$\Rs$& {[0.5, 2.7, 5.0, 7.3, 9.6, 11.8, 14.2, 16.4, 18.7, 21.0]}  &  \SI{}{\angstrom} \\
		$\Rc$& {[ 21.0]} &  \SI{}{\angstrom}  \\
		\hline
	\end{tabular}
\end{table}

\section{Additional sexithiophene structural distributions}

\begin{table}[H]
	\caption{\label{tab:table2}Optimal coarse-grained principal moments of inertia $I_q$ for $q = 1,2,3$, calculated using Eqs.~(9) and (10) of the main paper and the percentage difference between these values. }
	\centering
	\begin{tabular}{c | c | c | c}
	\hline
	 principal axis $q$ & $I_q$ (Eq.~(9)) (\si{\gram\per\mole\per\angstrom\squared})  & $I_q$ (Eq.~(10)) (\si{\gram\per\mole\per\angstrom\squared} )&\% difference \\
		\hline
		$1$& \num{1083.0} & \num{1080.8} & {0.2}  \\
		$2$& \num{20543.1}& \num{20712.5} &{0.8} \\
		$3$& \num{ 21280.8} & \num{21395.8} &{0.5}\\
		\hline
	\end{tabular}
\end{table}

\begin{figure}[H]
	\includegraphics{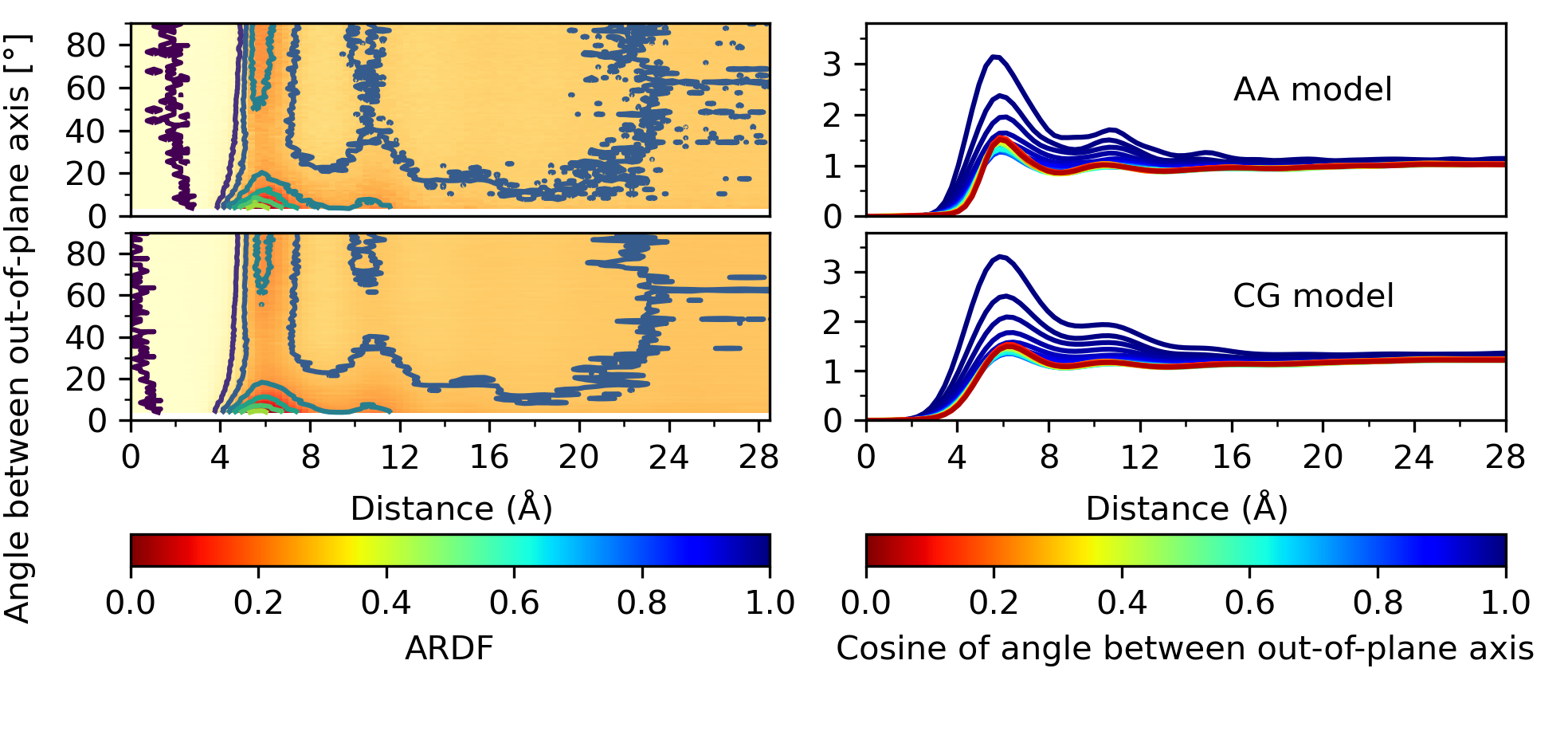}
	\caption{Angular--radial distribution function (ARDF) of the all-atom (AA) (top) and coarse-grained (CG) (bottom) sexithiophene models at 640~K and 1~atm depicted as a heat map (left) and 1D slices at constant angle (right). Face-to-face, edge-to-edge, and parallel displaced configurations occur when the angle is 0\textdegree, while edge-to-face configurations occur at 90\textdegree.}
	\label{fig:sexithiopheneArdf640}
\end{figure}

\section{TensorFlow and LAMMPS implementation requirements}

The following list of software is needed to train and use the neural network model in coarse-grained simulations.

\begin{enumerate}
	\item Tensorflow C API: https://github.com/tensorflow/tensorflow/blob/master/tensorflow/c/c\_api.h
	\item Cpp Flow: https://github.com/serizba/cppflow
	\item Tensorflow Python: https://github.com/tensorflow/tensorflow
	\item Keras: https://github.com/keras-team/keras
	\item LAMMPS: https://github.com/lammps/lammps
\end{enumerate}

The training and testing of the neural network potential was done with TensorFlow  in Python using the Keras functional API. The force and torque calculations were obtained through TensorFlow's Gradient Tape feature, which provides computational derivatives with respect to the network parameters. The tanh activation function was used for all standard neural network layers except the output layer since the tanh activation produced a smooth differentiable potential energy surface. The mean squared error was used when calculating the loss for the forces, torques, and virials. The Adam optimizer \cite{Kingma2014} was used as the gradient descent algorithm since it was able to reach the global minimum without manually updating the learning rate during training. The machine-learning potential was deployed with the TensorFlow C API and Cpp Flow wrapper. Cpp Flow allows the TensorFlow C model to be accessed directly as a force and torque calculator in a LAMMPS pair-style function.

\bibliography{afm-cg-nn}%